\documentclass[12pt]{article}

\newcommand{\blind}{0}

\usepackage{amsmath}
\usepackage{graphicx}
\usepackage{enumerate}
\usepackage{natbib} 
\usepackage{url} 

\graphicspath{{figures/}}
\usepackage{amsfonts, amsthm, latexsym, amssymb}
\usepackage{float}
\usepackage{multirow}
\usepackage{xcolor}
\usepackage{lineno}
\usepackage[ruled,vlined]{algorithm2e}
\newcommand{\mb}{\mathbf}
\newcommand{\mc}{\mathcal}
\newcommand{\bs}{\boldsymbol}

\begin{document}
	
	\def\spacingset#1{\renewcommand{\baselinestretch}%
		{#1}\small\normalsize} \spacingset{1}
	
	\if0\blind
	{
		\title{\bf Gaussian process aided function comparison using noisy scattered data}
		\author{Abhinav Prakash, Rui Tuo, and Yu Ding \\
			Department of Industrial and Systems Engineering\\
			Texas A\&M University}
		\date{}
		\maketitle
	} \fi
	
	\if1\blind
	{
		\bigskip
		\bigskip
		\bigskip
		\begin{center}
			{\LARGE\bf Gaussian process aided function comparison using noisy scattered data}
		\end{center}
		\medskip
	} \fi
	
	\bigskip
	\begin{abstract}
		This work proposes a nonparametric method to compare the underlying mean functions given two noisy datasets. The motivation for the work stems from an application of comparing wind turbine power curves. Comparing wind turbine data presents new problems, namely the need to identify the regions of difference in the input space and to quantify the extent of difference that is statistically significant. Our proposed method, referred to as funGP, estimates the underlying functions for different data samples using Gaussian process models. 
		We build a confidence band using the probability law of the estimated function differences under the null hypothesis.
		Then, the confidence band is used for the hypothesis test as well as for identifying the regions of difference. 
		This identification of difference regions is a distinct feature, as existing methods tend to conduct an overall hypothesis test stating whether two functions are different. Understanding the difference regions can lead to further practical insights and help devise better control and maintenance strategies for wind turbines. The merit of funGP is demonstrated by using three simulation studies and four real wind turbine datasets.
	\end{abstract}
	
	\noindent%
	
	\spacingset{2} 
	\section{Introduction}\label{Sec1}
	Comparing information from two datasets is an important topic in statistics. Various methods exist to compare datasets arising from univariate and multivariate distributions, for example, two sample $t$-test~\citep{fisher1925} and Hotelling's $T^2$ test~\citep{hotelling1931}, respectively. The literature is not just limited to comparing finite dimensional objects, but also extends to functions. In this article, we focus on nonparametric methods that compare functions.
	
	Our work is motivated by an application in the wind energy sector, where the goal is to compare two power curves. The power curve of a wind turbine is a function with wind power as the output and some environmental variables (such as wind speed, wind direction, air density) as the inputs. Power curves are used to characterize the performance of wind turbines~\citep{IEC05}. Hence, comparing power curves plays a critical role in assessing and benchmarking turbine performance, devising maintenance plans, and justifying expensive overhauls or retrofits~\citep{Hwangbo17,Ding2019}. Some important aspects for comparing power curves are to understand where the power curves differ (in the input space) and how much is the difference. 
	It is not only important to check whether two power curves differ, but more crucial to identify the regions of difference and quantify the difference for guiding economically justifiable actions.  In addition to the need for identifying the regions of difference, the datasets arising from wind turbines entails two other features: the input conditions (e.g. wind speed or wind direction) for the observations cannot be controlled, and as a result, the input points for any two datasets are not the same, and there lacks replicates for any input point. Taken altogether, our research objective is to to develop a nonparametric function comparison method that meets the following three requirements:
	\begin{itemize}
		\item The method can identify the regions between two functions that are statistically different and quantify the difference;
		\item The input data points associated with the two functions are not necessarily the same;
		\item There are no replications in the data points.
	\end{itemize}
	
	The problem of testing the equality of two nonparametric functions has been studied extensively in the literature. One early work is~\cite{HallHart1990}. They defined a test statistic for the problem using the smoothed (estimated) function values and obtained a distribution of their test statistic using bootstrap method. \cite{King1991}  also studied the same problem using smoothing techniques and proposed an exact distribution for their test statistic under the normality assumption for the errors. \cite{Delgado1993} proposed another test statistic using marked empirical process. \cite{FanLin98} worked on reducing the dimension of the problem using discrete Fourier transforms so that standard multivariate techniques can be used to test the hypothesis. These works assume that the two datasets under comparison have identical input points. This assumption is relaxed in~\cite{kulasekara1995,kulasekara1997,munk1998,neumeyer2003}, which propose tests that are valid under different input points among the datasets. The literature mentioned hitherto are \emph{global} tests, providing a binary answer on whether the functions are statistically the same or not. They do not provide any insights on the regions of the input space where the functions are different, or which function has higher or lower function values.
	
	\cite{CoxLee2008} addressed this problem of identifying difference regions, using a pointwise testing procedure based on the Westfall-Young randomization technique~\citep{westfall1993}. However, Cox and Lee's method does not meet our aforementioned requirements, because (a) \cite{CoxLee2008} requires replications of data points, as Cox and Lee's method is based on permutation, which requires data replication.  (b) \cite{CoxLee2008} identifies the region but cannot easily quantify the statistical difference.  Their method produces pointwise $p$-values rather than a coherent confidence band for functional differences.  It is not straightforward to convert the pointwise $p$-values into functional differences. (c) \cite{CoxLee2008} developed their method for the cases when the input points for the two functions are at the same locations.
	
	In this work, we propose a new nonparametric function comparison method that satisfies the three requirements posed above. We first use a Gaussian process (GP) regression model to recover the functions from the noisy datasets. Given a prescribed type I error, we then build a confidence band on the difference between the functions throughout the input space under the null hypothesis. If the actual difference between the functions computed using the data is beyond the confidence band, we reject the null hypothesis. We call the method \emph{function comparison using Gaussian Process} or \emph{\textbf{funGP}}.
	
	The funGP method does not require the input points among the samples to be the same, nor does it need replicates of the observations. When the null hypothesis is rejected, funGP identifies the regions of difference in the input space and quantifies the estimated difference using the established confidence band. Although we assume the functions as realizations of GPs, we demonstrate that the method works well for deterministic functions also. We apply our method to real wind turbine datasets and compare the results with some existing work for turbine performance characterization. That GP regression works for a large class of functions makes the proposed method applicable to many problems. GP models also provide uncertainty quantification, enabling a statistically reliable function comparison.
	
	We organize the rest of the paper as follows. Section 2 provides the details of the proposed method. Section 3 presents the simulation examples and comparison studies with two existing functional tests. We apply the funGP method to wind turbine datasets in Section 4. We conclude the work with some discussions in Section 5.
	
	\section{The funGP Method}\label{Sec2}
	In this section, we describe the mathematical formulation and the implementation details of the proposed funGP method.
	\subsection{Problem Formulation}\label{Sec2.1}
	Let us consider two datasets, $\{\mathcal{D}_i\ |\ i = 1,2\}$, with  $n_1$ and $n_2$ data points, respectively. Each data point consists of a $d$-dimensional input vector and a real-valued output. Assume that $ \mathcal{D}_1 $ can be denoted by an ordered pair $ \{\mathbf{X}^{(1)},\boldsymbol{y}^{(1)}\}$, where $ \mathbf{X}^{(1)} $ is a $ n_1 \times d $ matrix with each row corresponding to input variable values for one data point and $ \boldsymbol{y}^{(1)} $ is a vector of length $ n_1 $ with each component as response for one data point. Similarly $ \mathcal{D}_2 $ can be denoted as $ \{\mathbf{X}^{(2)},\boldsymbol{y}^{(2)} \} $. Specifically,
	\begin{equation*}
		\boldsymbol{y}^{(1)} =
		\begin{bmatrix}
			y_{11}\\
			y_{12}\\
			\vdots\\
			y_{1n_1}
		\end{bmatrix},
		\quad
		\mathbf{X}^{(1)} =
		\begin{bmatrix}
			-\boldsymbol{x}_{11}^\top-\\
			-\boldsymbol{x}_{12}^\top- \\
			\vdots\\
			-\boldsymbol{x}_{1n_{1}}^\top- \\
		\end{bmatrix},
		\quad
		\boldsymbol{y}^{(2)} =
		\begin{bmatrix}
			y_{21}\\
			y_{22}\\
			\vdots\\
			y_{2n_2}
		\end{bmatrix},
		\quad
		\mathbf{X}^{(2)} =
		\begin{bmatrix}
			-\boldsymbol{x}_{21}^\top- \\
			-\boldsymbol{x}_{22}^\top- \\
			\vdots\\
			-\boldsymbol{x}_{2n_{2}}^\top- \\
		\end{bmatrix}.
	\end{equation*}
	We also assume that these datasets come from underlying models given by:
	\begin{equation}\label{model}
		y_{ij} = f_i(\boldsymbol{x}_{ij}) + \epsilon_{ij}, \quad i = 1,2, \quad  j = 1,\dots,n_i,
	\end{equation}
	where $f_1(\cdot)$ and $f_2(\cdot)$ are  two smooth continuous functions with the same compact domain $\mathcal{X} \subset \mathbb{R}^d$ and $\epsilon_{ij} \stackrel{i.i.d.}{\sim} \mathcal{N}(0,\sigma_{\epsilon}^2)$ with a constant variance $ \sigma_{\epsilon}^2 < \infty$. We here consider the same noise variance for both datasets. This assumption is introduced only for simplicity and can be relaxed.
	
	The goal is to test the following null and alternative hypotheses. The null hypothesis is that the functions are identical, whereas the alternative hypothesis is that the functions differ for at least one $\boldsymbol{x} \in \mathcal{X}$.
	Under the null hypothesis, $H_0$:
	\begin{equation*}
		f_1(\boldsymbol{x}) = f_2(\boldsymbol{x}) \quad \text{for all }\ \boldsymbol{x} \in \mathcal{X}.
	\end{equation*}
	And, under the alternative hypothesis, $H_1$:
	\begin{equation*}
		\text{there exists } \quad \boldsymbol{x} \in \mathcal{X} \quad \text{such that } \quad f_1(\boldsymbol{x}) \neq f_2(\boldsymbol{x}).
	\end{equation*}
	
	A rigorous frequentist testing of the null hypothesis $H_0$ usually relies on a test statistic whose distribution is (approximately) independent of the underlying function $f:=f_1=f_2$ under $H_0$. One would consider using the estimator of $f_1-f_2$ to build a test statistic. Specifically, we in this work invoke a GP framework for calculating the distribution in the presence of an unknown $f$. This assumption allows for calculating the distribution of an intuitive estimator of $f_1-f_2$. Of course, doing this requires us to replace the null hypothesis $H_0$. Details will be presented in the next subsection.

 In addition to the above discussion, we stress that our application requires the test statistic to be a functional statistic, as we are interested in identifying the region of input space where the functions are different. Specifically, when $H_0$ is rejected, we need to identify the set $\mc{S} = \{\bs{x} : f_1(\bs{x}) \neq f_2(\bs{x}) \}$. Most of the existing methods reviewed in Section \ref{Sec1}, such as \cite{munk1998}, use a univariate statistic to test the hypothesis and cannot identify the region of difference. Under certain conditions, such as the input points of the two datasets are identical and replicated response, a functional test is available; see \cite{CoxLee2008}. To the best of our knowledge, no statistics have been proposed in the literature under the general conditions as in the current context.

	
	\subsection{Hypothesis testing with a GP prior}\label{Sec2.2}
	The general idea for a hypothesis testing is to find a test statistic and subsequently have a decision rule to either accept or reject the null hypothesis based on the value of the test statistic. In the application described, we are not only interested in the binary answer that whether the two functions are different, but also want to understand where the difference lies in the input space. This requires us to obtain a test statistic at the input points for which we do not have any data. We also assume that the input points for the two datasets are not the same. Thus, we would have to assume some structure in the functions (such as the functions are smooth and continuous) in order to recover the functions and estimate the noise in the model. Here we adopt a Bayesian idea that imposes a prior structure on the functions. Specifically, we use a GP prior with zero mean and a covariance function given by $ k(\boldsymbol{x},\boldsymbol{x}')$. The zero mean assumption is for mathematical simplicity, and we can assume a different mean function, if necessary.
	
	Despite the use of the GP prior, we still follow a frequentist hypothesis testing framework, by considering a new null hypothesis $H_0^{GP}$,  still stating $	 f_1(\boldsymbol{x}) = f_2(\boldsymbol{x}) \, \text{for all}\ \boldsymbol{x} \in \mathcal{X}$, but incorporating the following prior information:
	\begin{eqnarray*}
		y_{ij}&=&f(x_{ij})+\epsilon_{ij},\\
		f&\sim& \mathcal{GP}(0,k(\bf{x},\bf{x}')),
	\end{eqnarray*}
	where $\epsilon_{ij} \stackrel{i.i.d.}{\sim} \mathcal{N}(0,\sigma_{\epsilon}^2)$. In Sections~\ref{Sec2.2} and~\ref{Sec2.3}, we assume that $k(\cdot,\cdot)$ and $\sigma_{\epsilon}^2$ are known.
	
	So far, our goal can be described as testing $H_0^{GP}$ against the alternative hypothesis $H_1$. We will propose a test method, so that its type-I error under $H_0^{GP}$ has a probability controlled by a prespecified significance level $\alpha$. Note that the type-I error under $H_0^{GP}$ is
	$$\int_f \mathbb{P}(H_0^{GP}\text{ is rejected}|f_1=f_2=f) d \mathbb{P}_{GP}, $$
	where $\mathbb{P}_{GP}$ denotes the probability measure of the GP prior. It is worth noting that the type-I above is not identical to the type-I error under the original null hypothesis $H_0$. However, we expect that the proposed method can serve as an approximate method for the fixed-function testing problems, and we will verify this expectation via numerical studies in Section \ref{Sec3}.
	
	The main idea of our test is as follows. First we can reconstruct $f$ based on the datasets $\mathcal{D}_1$ and $\mathcal{D}_2$ separately, and denote the reconstructed functions as $\hat{f}_1$ and $\hat{f}_2$, respectively. Under $H_0^{GP}$, $\hat{f}_1$ and $\hat{f}_2$ should be close. Thus we can test $H_0^{GP}$ by computing the difference between $\hat{f}_1$ and $\hat{f}_2$.

	
	
	To reconstruct $f$, We start off by defining a cross-covariance matrix $ \mathbf{K_{X, X'}} $ between a pair of input variable matrix $ \mathbf{X} $ and $ \mathbf{X}' $, and a covariance vector $ \boldsymbol{r}(\boldsymbol{x}) $ between the input data $ \mathbf{X} $ and any point $ \boldsymbol{x} $ as follows:
	\begin{eqnarray}\label{CovMatNotations}
		\mathbf{K_{X, X'}} =
		\begin{bmatrix}
			k(\boldsymbol{x}_1,\boldsymbol{x}'_1) & k(\boldsymbol{x}_1,\boldsymbol{x}'_2) & \dots & k(\boldsymbol{x}_1,\boldsymbol{x}'_n)\\
			k(\boldsymbol{x}_2,\boldsymbol{x}'_1) & k(\boldsymbol{x}_2,\boldsymbol{x}'_2) &\dots  & k(\boldsymbol{x}_2,\boldsymbol{x}'_n)\\
			\vdots & \vdots & \ddots & \vdots \\
			k(\boldsymbol{x}_m,\boldsymbol{x}'_1) & k(\boldsymbol{x}_m,\boldsymbol{x}'_2) & \dots & k(\boldsymbol{x}_m,\boldsymbol{x}'_n)\\
		\end{bmatrix},
		\quad
		\boldsymbol{r}(\boldsymbol{x}) =
		\begin{bmatrix}
			k(\boldsymbol{x}_1,\boldsymbol{x})\\
			k(\boldsymbol{x}_2,\boldsymbol{x})\\
			\vdots\\
			k(\boldsymbol{x}_m,\boldsymbol{x})\\
		\end{bmatrix},
	\end{eqnarray}
	where $ \boldsymbol{x}_1 \dots \boldsymbol{x}_m $ are the vectors in the rows of the matrix $ \mathbf{X} $, and $ \boldsymbol{x}'_1 \dots \boldsymbol{x}'_n $ are the vectors in the rows of the matrix $ \mathbf{X}'$. When  $ \mathbf{X}= \mathbf{X}'$, $\mathbf {K}_{\mb{X},\mb{X}}$ is then a symmetric covariance matrix.
	The standard GP prediction theory suggests~\citep{Rasmussen2006}
	\begin{equation}\label{posteroirFn1}
		\hat{f}_1(\boldsymbol{x}) = \boldsymbol{r}_1(\boldsymbol{x})^\top[\mathbf{K}_{\mb{X}^{(1)},\mb{X}^{(1)}}+\sigma_{\epsilon}^2\mathbf{I}_{n_1}]^{-1}\boldsymbol{y}^{(1)},
	\end{equation}
	\begin{equation}\label{posteroirFn2}
		\hat{f}_2(\boldsymbol{x}) = \boldsymbol{r}_2(\boldsymbol{x})^\top[\mathbf{K}_{\mb{X}^{(2)},\mb{X}^{(2)}}+\sigma_{\epsilon}^2\mathbf{I}_{n_2}]^{-1}\boldsymbol{y}^{(2)},
	\end{equation}
	where $ \boldsymbol{r}_1(\boldsymbol{x}) $ is the covariance vector between $ \mathbf{X}^{(1)} $ and any point $ \boldsymbol{x} $, $ \mathbf{K}_{\mb{X}^{(1)},\mb{X}^{(1)}}$ is the covariance matrix for $ \mathbf{X}^{(1)} $, and $ \mathbf{I}_{n_1} $ is the identity matrix of proper size---$ n_1 \times n_1$ in this case. The notations in Equation \eqref{posteroirFn2} are likewise defined.
	
	
	%
	It is worth noting that although $\hat{f}_1(\bs{x})$ and $\hat{f}_2(\bs{x})$ are posterior means from a Bayesian perspective, here we take a frequentist point of view and regard them merely as statistics, i.e., functions of the data. To test the null hypothesis $H_0^{GP}$, we use the statistic $G(\bs{x}):=\hat{f}_1(\bs{x})-\hat{f}_2(\bs{x})$. Clearly, given $(\bs{X}^{(1)},\bs{X}^{(2)})^T$, the randomness of $G(\bs{x})$ comes solely from the data $(\bs{y}^{(1)},\bs{y}^{(2)})^T$, which follows a zero-mean multivariate normal distribution under $H_0^{GP}$. Therefore,  under $H_0^{GP}$ and given the input data, $G(\bs{x})$ is a centered GP
	%
 and  we write
	\begin{equation}\label{statistic}
		G(\cdot)|\bs{X}^{(1)},\bs{X}^{(2)},H_0^{GP} \sim  \mathcal{GP}(0,c(\cdot,\cdot)).
	\end{equation}
	After some calculations, we derive the expression for covariance function $ c(\cdot,\cdot)$ (see Appendix A.1 for details), given by
	\begin{equation}\label{DiffCov}
		\begin{split}
			&c(\bs{x},\bs{x}')  =  \mathbf{r}_2(\bs{x})^\top[\mathbf{K}_{\mb{X}^{(2)},\mb{X}^{(2)}}+\sigma_{\epsilon}^2\mathbf{I}_{n_2}]^{-1} \mathbf{r}_2(\bs{x}')
			+ \mathbf{r}_1(\bs{x})^\top[\mathbf{K}_{\mb{X}^{(1)},\mb{X}^{(1)}}+\sigma_{\epsilon}^2\mathbf{I}_{n_1}]^{-1}\mathbf{r}_1(\bs{x}')\\
			& - 2\mathbf{r}_2(\bs{x})^\top[\mathbf{K}_{\mb{X}^{(2)},\mb{X}^{(2)}}+\sigma_{\epsilon}^2\mathbf{I}_{n_2}]^{-1} \mathbf{K}_{\mb{X}^{(2)},\mb{X}^{(1)}} \ [\mathbf{K}_{\mb{X}^{(1)},\mb{X}^{(1)}}+\sigma_{\epsilon}^2\mathbf{I}_{n_1}]^{-1}\mathbf{r}_1(\bs{x}').
		\end{split}
	\end{equation}
	
	
	Therefore, to test $H_0^{GP}$, it suffices to find a $1-\alpha$ probability band of $ G(\cdot) $ under $H_0^{GP}$, i.e, a pair of functions $l(\bs{x})$ and $u(\bs{x})$ such that
	$$\mathbb{P}_{G\sim \mathcal{GP}(0,c(\cdot,\cdot))}\big(l(\bs{x})\leq G(\bs{x})\leq u(\bs{x}) \text{ for all } \bs{x}\big)\geq 1-\alpha. $$
	With a slight abuse of terminology, we shall call the band between $l(\bs{x})$ and $u(\bs{x})$ a $1-\alpha$ \textit{confidence band} for $\mathcal{GP}(0,c(\cdot,\cdot))$. It is worth noting that this band is related to the distribution of $G(\bs{x})$ only under the null hypothesis $H_0^{GP}$. The test then proceeds by checking whether $ G(\cdot) $ remains within the confidence band. If there exists an $\boldsymbol{x}$ for which $ G(\bs{x}) $ is outside the band then we reject the null hypothesis. Clearly, such a testing method ensures a $1-\alpha$ type-I error under $H_0^{GP}$. The question now is how to efficiently build an effective confidence band for $\mathcal{GP}(0,c(\cdot,\cdot))$.
	
	\subsection{Building the confidence band}\label{Sec2.3}
	To build a $1-\alpha$ confidence band for $\mathcal{GP}(0,c(\cdot,\cdot))$, the main idea is to sample from a set with a coverage probability of $1-\alpha$. For notational simplicity, we suppose $G(\bs{x})\sim \mathcal{GP}(0,c(\cdot,\cdot))$ in this subsection, i.e., the null hypothesis $H_0^{GP}$ is true. 
	Since $G(\bs{x})$ is an infinite dimensional object, it is more convenient to work with a finite dimensional representation of it.
	To this end, we employ the Karhunen Lo\`{e}ve (KL) expansion on $G(\bs{x})$. The KL expansion for the zero mean Gaussian process $G(\bs{x})$ is given as follows:
	$G(\boldsymbol{x}) = \sum_{k=1}^{\infty}\sqrt{\lambda_k} \phi_k(\boldsymbol{x}) z_k$,
	where $\{z_k\}_{k = 1}^\infty$ are uncorrelated standard normal random variables,
	$\{\phi_k(\cdot)\}_{k = 1}^\infty$ are the basis functions,  $\{\lambda_k\}_{k = 1}^\infty$ are the eigenvalues. In practice, this infinite sum is truncated for two reasons: 1) If the process is smooth, the eigenvalues would decay rapidly, 2) To make the computation tractable. Under the assumption that the underlying functions under comparison are smooth, we discard all the eigenvalues smaller than a certain threshold. For practical purposes, we find a threshold value of $10^{-6}\times \lambda_{max}$ to work well, where $\lambda_{max}$ is the largest eigenvalue. Thus, the KL expansion decomposes the process into independent components and also reduces the dimension of the problem by finding a sparse representation of the process using its eigenfunction basis.
	
	Let the truncation number computed from the aforementioned rule be $m$, that is, only the $m$ largest eigenvalues are significantly ``large". Then, we write the truncated KL expansion for $G(\bs{x})$ as follows:
	\begin{equation}\label{KLHT}
		G(\bs{x}) \approx \sum_{k=1}^{m}\sqrt{\lambda_k} \phi_k(\boldsymbol{x}) z_k
	\end{equation}
	In Equation \eqref{KLHT}, the randomness is introduced by $z_k$'s. Hence, in order to build a $1-\alpha$ confidence band on $G(\bs{x})$, we build the same level confidence band for the joint distribution of $z_k \ | \ k = 1,\dots,m$, which can be constructed as follows. We know that for an $m$-dimensional uncorrelated standard normal vector, a confidence region $\mathcal{R}$, with probability $\mathbb{P}(\mathcal{R}) = 1-\alpha$, can be described using a hypersphere of radius $r$. This radius can be expressed as $r = ||\boldsymbol{z}||$, where $||\boldsymbol{z}|| = ||z_1^2 + z_2^2 + \cdots + z_{m}^2||$ is the $\ell^2$ norm. The sum of squares of $m$ uncorrelated standard normals follows a chi-square distribution with $m$ degrees of freedom, that is,
	$
	r^2 = z_1^2 + z_2^2 + \cdots + z_{m}^2 \sim \chi^2_{m}.
	$
	Hence, $r$ is computed by inverting the CDF of a chi-square distribution in the following way:
	\begin{equation}\label{Eqn:radius}
		r = \sqrt{F_{m}^{-1}(1-\alpha)},
	\end{equation}
	where $F_{m}^{-1}(\cdot)$ is the inverse CDF of $\chi^2_{m}$. Once we have the radius $r$, we sample $\boldsymbol{z}$ from the region with a coverage probability of $1-\alpha$ using the following rule:
	\begin{itemize}
		\item sample $ z_i $ from $\mathcal{N}(0,1) \ | \ i \in \{ 1 \dots m  \}$,
		\item if  $\sum_{i=1}^{m} z_i^2 \leq r^2$, accept $ \boldsymbol{z} = (z_1, z_2, \dots, z_{m})^\top $; else, reject it.
	\end{itemize}
	The samples of $\bs{z}$ obtained as above can be easily converted to samples from $G(\bs{x})$ that are from $1-\alpha$ confidence set using Equation \eqref{KLHT}. In order to test the hypothesis, we would need to compare the actual difference in the predictive means $g(\bs{x})$ with the confidence band at all the points $\bs{x} \in \mathcal{X}$. This is practically intractable, as for any continuous function, there are infinitely many points in the domain. Hence, we discretize the domain using a finite-sized evenly spaced test grid to approximate $\mathcal{X}$. Let $ \mathbf{X}_{test} $ be a $n_{test} \times d$ matrix with each row corresponding to one grid point $\bs{x}_{t_j} \ |\ j = 1,\dots,n_{test} $. We compare the function difference with the confidence band on these grid points. Testing on this regular grid is a reasonable approximation to testing for all $ \boldsymbol{x} \in \mathcal{X}$ because of our underlying assumption that the functions are continuous and smooth. Let $\mb C_{\mb X_{test}, \mb X_{test}}$ be the covariance matrix generated using the covariance function $c(\bs{x},\bs{x}')$ using all the points in $\mb{X}_{test}$ in a similar way as $\mb{K}_{\mb{X},\mb{X}'}$ is defined in Equation \eqref{CovMatNotations}. Let $\mb{\Lambda}$ be an $m \times m$  diagonal matrix with $m$ largest eigenvalues of $\mb{C}_{\mb{X}_{test},\mb{X}_{test}}$ and let $\mb{U}$ be an $n_{test} \times m$ matrix whose columns are the eigenvectors corresponding to the $m$ largest eigenvalues of $\mb{C}_{\mb{X}_{test},\mb{X}_{test}}$. Then, the KL expansion at all the points in $\mb{X}_{test}$, denoted by a random vector $\mb{G}$ such that its $j^{th}$ component $(\mb{G})_j = G(\bs{x}_{t_j})$, can be expressed using the matrix notation as follows (see Appendix A.2 for details):
	\begin{equation}
		\mb{G}  = {\mathbf{U}}{\mathbf{\Lambda}}^{\frac{1}{2}}{\boldsymbol{z}}.
	\end{equation}

	In order to construct the confidence band, we sample a large number (say 1,000) of $\boldsymbol{z}$ from its confidence set, then the values of the confidence band at all points in $\mb{X}_{test}$ is given by the vectors:
	\begin{equation}\label{Eqn:band}
		\begin{split}
			\boldsymbol{ub} & = \text{Max}_{\boldsymbol{z}}  {\mathbf{U}}{\mathbf{\Lambda}}^{\frac{1}{2}}{\boldsymbol{z}}, \\
			\boldsymbol{lb} & = -\boldsymbol{ub}, \\
		\end{split}
	\end{equation}
	where $\boldsymbol{ub}$ is the vector of upper bounds and $\boldsymbol{lb}$ is the vector of lower bounds for the confidence band. We accept the null hypothesis $H_0^{GP}$ at the confidence level of $1-\alpha$ if the value of $g(\bs{x}_{t_j})$ are within the band, that is,
	\begin{equation*}
		(\boldsymbol{lb})_j \leq g(\bs{x}_{t_j}) \leq (\boldsymbol{ub})_j \quad \text{ for all } \ j = 1\dots n_{test},
	\end{equation*}
	where the notation $(\textbf{a})_j$ is the $j^{th}$ component of a vector $\textbf{a}$. Similarly, we reject the null hypothesis $H_0^{GP}$ at the $1-\alpha$ confidence level if there is at least one violation, that is,
	\begin{equation*}
		\text{there exists } \ j \in \{1\dots n_{test}\}\ \text{such that } \ g(\bs{x}_{t_j}) \notin [(\boldsymbol{lb})_j,  (\boldsymbol{ub})_j ].
	\end{equation*}	
	
	It is also worth noting that we are getting an approximate band because of using a truncated KL expansion. This may result in losing some confidence on the test.  In other words, the probability of the confidence band would be less than $1-\alpha$. A possible compensation can be made by setting a slightly higher confidence level than the nominal level.
	
	The points for which the null hypothesis is rejected would form a discrete grid on the difference region(s) and the absolute value of the statistically significant difference at these points would be given as:
	$\delta(\bs{x}_{t_j}) = |g(\bs{x}_{t_j})| - (\bs{ub})_{j}$. Needless to say that the difference at the points where the null hypothesis is not rejected would be considered zero.

	\subsection{Estimating the hyperparameters}\label{Sec2.4}
	Until now, we have assumed the values of the hyperparameters of the covariance matrix and the nugget $ \sigma_{\epsilon} $ are known. Next, we describe the method we use to estimate these hyperparameters.
	
	Let us assume that $\boldsymbol{\theta}$ is the vector containing all the hyperparameters of the covariance function and the nugget $ \sigma_{\epsilon} $. We estimate these hyperparameters by merging the two datasets as $\bs{y} = (\bs{y}^{(1)}, \bs{y}^{(2)})$ and $\mb{X} = (\mb{X}^{(1)},\mb{X}^{(2)})$, and jointly maximizing the likelihood as follows:
	\begin{equation}\label{Eqn:Likelihood}
		\boldsymbol{\hat{\theta}} = \arg\max\ \mathcal{L}(\boldsymbol{\theta};\mathcal{D}_1,\mathcal{D}_2),
	\end{equation}
	where $\mathcal{L}(\boldsymbol{\theta};\mathcal{D}_1,\mathcal{D}_2) = \frac{1}{(2\pi)^{(n_1+n_2/2)}\lvert \mb{K}_{\mb{X},\mb{X}}+ \sigma_{\epsilon}^2 \mb{I} \rvert} e^{(-\bs{y}^\top [\mb{K}_{\mb{X},\mb{X}}+ \sigma_{\epsilon}^2\mb{I}]^{-1} \bs{y})}$.
	We provide a summary of the funGP algorithm as Appendix A.3.
	
	\section{Simulation Study}\label{Sec3}
	In this section, we present three simulation studies for the funGP method to estimate the type I and the type II errors and compare it with two methods from the existing literature. We estimate the type II error for some small perturbations. In order to quantify the difference between a function and its perturbation, we use an $L^2$-distance percentage defined as follows:
	\begin{equation*}
		L^2 \ dist \ \% =  \frac{\lVert f - g\rVert_{L^2}}{\lVert f\rVert_{L^2}} \times 100\%,
	\end{equation*}
	where $ f $ is the underlying function, and $ g  $ is its perturbation. After fixing the nominal level of $H_0^{GP}$ to $\alpha = 0.05$, we conduct 1,000 runs for each simulation example to estimate the type I/type II errors. We also examine the effect of the sample size, the number of points in the test grid, and the truncation number in the KL expansion on the type I and type II errors using different experiments.
	
	\subsection{Functions used in the simulations}\label{Sec3.1}
	The first simulation study is based on functions sampled from a known Gaussian process. Our method also assumes the functions to be GP samples, so this study represents a case when there is no model misspecification, that is, the $GP$ part in $H_0^{GP}$ is indeed true. Whereas the other two simulation studies are based on some parametric functions available in the literature; we use GP as a surrogate for the true function. Thus, for these two fixed functions, the estimated type I/type II errors are for the original hypothesis test $H_0$, even though we are controlling the type I error only under $H_0^{GP}$. Hence, the last two simulations evaluate the efficacy of the funGP method under a potential model misspecification. In order to generate the datasets, we randomly sample two sets of points from the input domain of the functions. We then generate response by adding some i.i.d Gaussian noise to the function values at the sampled input points. For conducting all the simulation studies, we use a constant mean and a squared exponential covariance function for the GP modeling.
	
\begin{figure}
		\centering
		\includegraphics[width=2.5in]{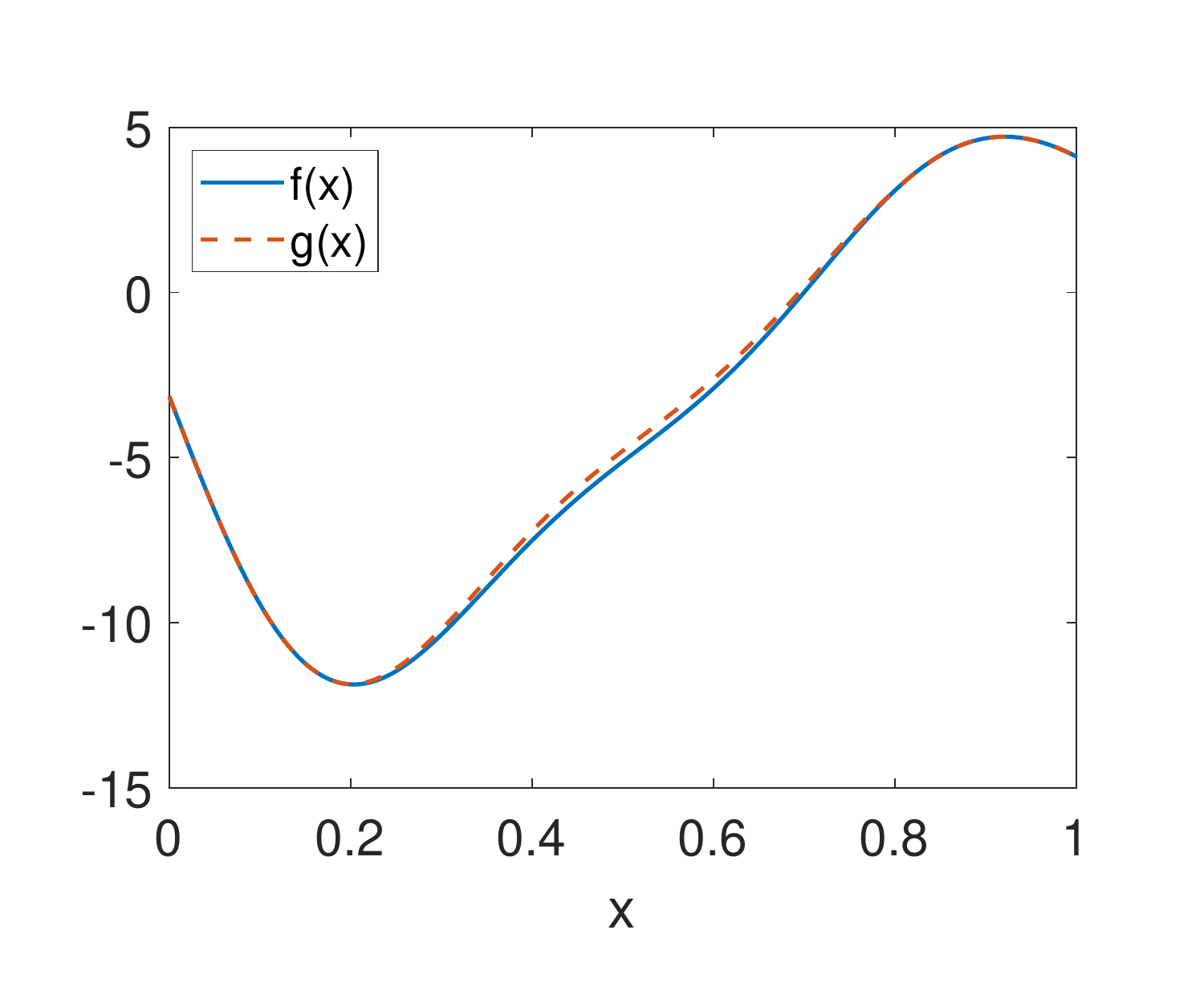}\qquad
		\includegraphics[width=2.5in]{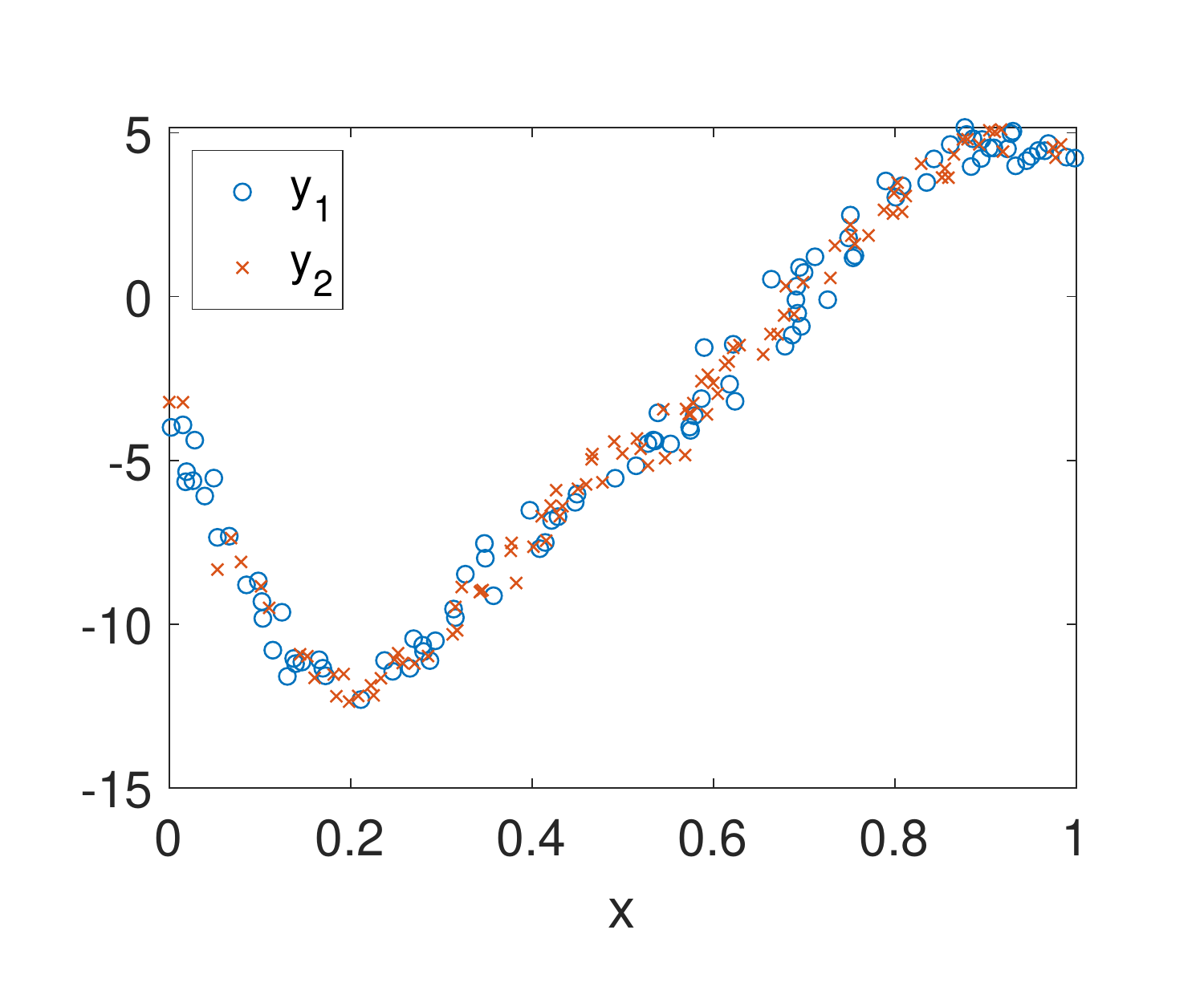}
		\caption{Plots for GP sample. Left panel: $ f(x) $ and its perturbation, $ g(x) $; Right panel: Noisy realizations from $ f(x) $ and $ g(x) $.}\label{GPsamplePlot}
	\end{figure}

	The following are the specifications of the simulated functions. We consider a one-dimensional input $x \in [0,1]$ for the first simulation study, the GP sample. The model can be described as:
	$y =  f(x) + \epsilon; \ f(x)  \sim \mathcal{GP}(0,k(x,x')); \ \epsilon  \sim \mathcal{N}(0,\sigma_{\epsilon}^2)$.
	The covariance function $k(x,x')$ is squared exponential with the following form:
	$k(x,x') = \sigma_f^2 \ \text{exp}\Big(- \Big[\frac{x-x'}{\theta}\Big]^2\Big)$. The hyperparameters for the covariance function, $ k(x,x') $, are set to $ \sigma_f = 5 $, and $ \theta = 0.2 $. The standard deviation of the noise, $ \sigma_{\epsilon} $, is set to $ 0.5 $. For each simulation run, a different sample is generated from the given GP model, and the estimated type I error is the percentage of runs for which the null hypothesis is rejected. For estimating the type II error, we create a perturbation $ g(x) $ in the following way:
	\[
	g(x) = \begin{cases}
	f(x)  + \frac{1}{3} \sin\Big(\pi\Big(\frac{x-0.2}{0.8 - 0.2}\Big)\Big), & x \in [0.2,0.8],\\
	f(x), & \text{otherwise.}
	\end{cases}
	\]
	The functions $ f(x) $ and $ g(x) $ sampled for one simulation run (left panel) along with two noisy datasets generated from it (right panel) are shown in Figure~\ref{GPsamplePlot}. One can see that the difference between the functions is small and gets masked visually in the noisy data.
	
	For the other two studies, we use two parametric functions available in the literature: piston simulation function~\citep{Kenett_Zacks_1998} and borehole simulation function~\citep{harper1983}. We use these functions with two dimensional input by fixing the rest of their input variables to certain values. More details about these functions, including the function plots, are provided as Appendix A.4 to maintain the flow for the readers and save space.
	
	\subsection{Results}\label{Sec3.2}
	Table~\ref{Table:SimulationResults} shows the estimated type I and type II errors for all the simulation studies, along with the $ L^2 $ distance between the function $ f $ and its perturbation, $ g $. The results in Table~\ref{Table:SimulationResults} are based on the following specifications. The sample size for one-dimensional function (GP sample) and two-dimensional functions (piston and borehole) are 500 and 1,000, respectively, randomly sampled from their respective input domain. The test grid is 500 evenly spaced points in the domain for the GP sample and 50 $\times$ 50 evenly spaced grid for the piston and the borehole functions. We follow the same truncation rule for the truncation number $m$ as described in Section~\ref{Sec2.3}. The $ L^2 $ distance varies between 3 to 5 \%.
	
	\begin{table}[h]
		\spacingset{1}
		\caption{Estimated type I and type II errors for the simulated functions.}\label{Table:SimulationResults}
		\vspace{6 pt}
		\centering
		\begin{tabular}{c| c  c | c  }
			\hline
			\textbf{Function} & Type I error & Type II error & $ L^2 \ dist \ \% $\\
			\hline
			GP sample & 0.049 & 0.031 & 4.7 \\
			Piston & 0.041 & 0.008 & 3.8\\
			Borehole  & 0.065 & 0.022 & 3.4 \\
			\hline			
		\end{tabular}
	\end{table}
	
	For the first case study, when the true functions are GP samples, the estimated type I error is very close to the nominal level of 0.05 (5\%). This result is very much expected, as both we are controlling the type I error under $H_0^{GP}$ and the GP assumption is indeed true. In the other two simulation studies, the estimated type I error is not as close to the nominal value as the first simulation study. This can be attributed to the fact that the estimated type I error is for $H_0$, and we are controlling the type I error for $H_0^{GP}$. The form of the mean and covariance function required to sample these functions from a GP is not known and we use approximations in these studies. The agreement between the estimated type I error and the nominal value would depend on how well the GP approximates the function. If it is difficult to approximate a function using a known parametric covariance function, we can either come up with more sophisticated mean and covariance functions, or we can increase the confidence level of the test to a value greater than the desired level to account for model uncertainty. For the given sample size, we are satisfied that the method can identify the difference in the underlying functions even with small perturbations. We would, next, conduct experiments to see how the method performs under different sample sizes, test grid sizes and truncation numbers.
	
	\subsection{Further experiments}\label{Sec3.3}
	We repeat the three simulation experiments carried out previously under different sample sizes while keeping the test grid and the truncation rule fixed. The sample sizes are set at four levels: 100, 200, 500, and 1,000.
	Table~\ref{sampleSizeExperiment} presents the results of these experiments. The table
	clearly shows a reduction in type II error as the sample size increases while keeping the type I error stable, which is consistent with our understanding of statistical hypothesis tests.
	The numerical results indicate that in order to render sufficient detection power, a large enough sample is needed for detecting small difference between two functions.
	
	\begin{table}
		\spacingset{1}
		\caption{Estimated type I and type II errors under different sample sizes.}\label{sampleSizeExperiment}
		\vspace{6 pt}
		\centering
		\begin{tabular}{c|c | c c c c }
			\hline
			\multirow{2}{5em}{\textbf{Function}} & \multirow{2}{4em}{Estimate} & \multicolumn{4}{c}{Sample size for each dataset}\\ \cline{3-6}
			& & 100 & 200 & 500 & 1,000 \\
			\hline
			\multirow{2}{4em}{GP sample} & Type I error & 0.032 & 0.040 & 0.049 & 0.049 \\
			& Type II error & 0.721 & 0.450 & 0.031 & 0.001\\
			\hline
			\multirow{2}{4em}{piston} & Type I error & 0.031 & 0.032 & 0.027 & 0.041\\
			& Type II error & 0.845 & 0.639 & 0.201 & 0.008\\
			\hline
			\multirow{2}{4em}{borehole} & Type I error & 0.058 & 0.058 & 0.070 & 0.065\\
			& Type II error & 0.757 & 0.460 & 0.091 & 0.022\\
			\hline
		\end{tabular}
	\end{table}
	
	The test grid size experiment is carried out while keeping the sample size fixed at the same value as used for the main result in Section~\ref{Sec3.2}. We use a test grid of size 100, 400, 900, and 2,500 for each simulated function so that it corresponds to 10$\times$10, 20$\times$20, 30$\times$30, and 50$\times$50 test grid, respectively, for two-dimensional functions (piston and borehole). Table~\ref{testGridSizeExperiment} presents the results of this experiment. There is no significant effect of the test grid size on the type I and type II errors. This is expected as our test relies on the truncated KL expansion and the number of eigenvalues ($m$) remains constant for different test grid sizes because we use the same truncation rule.
	
	We contemplate how the results may change when we use any arbitrary truncation number instead of using the aforementioned rule for calculating the truncation number. Table~\ref{truncNumberExperiment} displays the result of using different truncation numbers on the type I and type II errors. We note that the hypothesis test remains a level-$\alpha$ test as long as the truncation number is larger than a certain threshold, of which the specific value would depend on the function under study. For the first two cases, a truncation number of 10 or greater appears sufficient, whereas for the third case, a truncation number may need to be as large as 50. When one chooses a smaller truncation number than the problem demands, then one cuts off a significant portion of the $1-\alpha$ confidence band, resulting in a high type I error. Thus, we suggest using the recommended truncation rule, which adapts the truncation number according to the problem.
	
	\begin{table}
		\spacingset{1}
		\caption{Estimated type I and type II errors under different test grid sizes.}\label{testGridSizeExperiment}
		\vspace{6 pt}
		\centering
		\begin{tabular}{c|c | c c c c }
			\hline
			\multirow{2}{5em}{\textbf{Function}} & \multirow{2}{4em}{Estimate} & \multicolumn{4}{c}{Number of test points}\\ \cline{3-6}
			& & 100 & 400 & 900 & 2500 \\
			\hline
			\multirow{2}{4em}{GP sample} & Type I error & 0.039 & 0.038 & 0.059 & 0.049 \\
			& Type II error & 0.033 & 0.033 & 0.034 & 0.022\\
			\hline
			\multirow{2}{4em}{piston} & Type I error & 0.025 & 0.032 & 0.044 & 0.041\\
			& Type II error & 0.020 & 0.014 & 0.014 & 0.008\\
			\hline
			\multirow{2}{4em}{borehole} & Type I error & 0.035 & 0.050 & 0.046 & 0.065\\
			& Type II error & 0.063 & 0.027 & 0.024 & 0.022\\
			\hline
		\end{tabular}
	\end{table}
	\begin{table}
		\spacingset{1}
		\caption{Estimated type I and type II errors under different truncation numbers.}\label{truncNumberExperiment}
		\vspace{6 pt}
		\centering
		\begin{tabular}{c|c | c c c }
			\hline
			\multirow{2}{5em}{\textbf{Function}} & \multirow{2}{4em}{Estimate} & \multicolumn{3}{c}{Truncation number}\\ \cline{3-5}
			& & 10 & 50 & 100 \\
			\hline
			\multirow{2}{4em}{GP sample} & Type I error & 0.051 & 0.034 & 0.029  \\
			& Type II error & 0.023 & 0.043 & 0.048 \\
			\hline
			\multirow{2}{4em}{piston} & Type I error & 0.036 & 0.013 & 0.024 \\
			& Type II error & 0.016 & 0.020 & 0.026 \\
			\hline
			\multirow{2}{4em}{borehole} & Type I error & 0.260 & 0.046 & 0.039 \\
			& Type II error & 0.014 & 0.020 & 0.018 \\
			\hline
		\end{tabular}
	\end{table}
	\subsection{Comparison with other methods}\label{Sec3.4}
	We compare our method with two other methods available in the literature. The first comparison is with~\cite{munk1998}, which is a global test that works for datasets without requiring common input points and replicates. Although it is a global test, we can still use this method to compare with the funGP method in terms of the type I and type II errors. This method builds its test statistic based on the $L^2$-distance between the functions. The method is developed for functions with one-dimensional input. For this reason, we use this method only for the first simulation study, the GP sample.
	
	Table~\ref{MD_compStudy} presents the results for the comparison. We note that out method is significantly more powerful than Munk and Dette's method. \cite{munk1998} provided an expression for approximating the power of their test, given the $L^2$-distance between the functions, the sample size, and the noise level; see Equation (17) in~\citet{munk1998}. The approximate power computed using that expression is 0.435, which is consistent with the empirically estimated type II error in Table~\ref{MD_compStudy} (power = $1 -$type II error).
	\begin{table}[h]
		\spacingset{1}
		\caption{Comparison between the funGP and Munk \& Dette (1998) methods for the GP sample simulation study.}\label{MD_compStudy}
		\vspace{6 pt}
		\centering
		\begin{tabular}{c|c| c c}
			\hline
			Function & Method &Type I error & Type II error \\
			\hline
			\multirow{2}{4em}{GP sample} & funGP & 0.049 & 0.031 \\
			& Munk \& Dette & 0.117 & 0.570 \\
			\hline
		\end{tabular}
	\end{table}
	
	We also compare our method with~\cite{CoxLee2008}, which identifies the difference region in terms of $p$-values. \cite{CoxLee2008} is based on a permutation test and requires the datasets to have replicates and the same input points. Since the datasets simulated for the funGP method do not have replicates and do not share the same input points, we simulate different sets of samples with replicates keeping the input points the same for the two functions. We apply both funGP and Cox and Lee's methods to these newly generated datasets to estimate the type I and type II errors. We still use 1,000 runs for the simulation. We use 50 input points with 10 replications each for 1-dimensional case (GP sample function) and 100 input points with 10 replications each for 2-dimensional cases (piston and borehole functions). The sample sizes are chosen such that the total number of the samples is equal to that of the main simulation study, that is, 500 for 1-dimensional case and 1,000 for 2-dimensional case. In each case, the nominal level of the test is set to $\alpha = 0.05$.
	
	The results for this comparison are presented in Table~\ref{CL_comp}.
	The proposed funGP method performs better than Cox and Lee in two out of three cases---the GP sample and the piston cases, and worse for the borehole case, in terms of the type II error. We would like to articulate that we advocate the merit of our method as identifying the difference region and quantifying the difference for datasets arising from a broader setting, namely without the same input points and replicates, and not purely in terms of its power in a binary decision. Yet, our method performs comparably, and sometimes even better, than other methods such as~\cite{CoxLee2008} and~\cite{munk1998}.

	\begin{table}
		\spacingset{1}
		\caption{Comparison between funGP and Cox and Lee methods.}\label{CL_comp}
		\centering
		\begin{tabular}{c|c| c c}
			\hline
			Function & Method &Type I error & Type II error \\
			\hline
			\multirow{2}{4em}{GP sample} & funGP & 0.039 & 0.042\\
			& Cox \& Lee & 0.023 & 0.160 \\
			
			\hline
			\multirow{2}{4em}{piston} & funGP & 0.027 & 0.011\\
			& Cox \& Lee & 0.017 & 0.071 \\
			
			\hline
			\multirow{2}{4em}{borehole} & funGP & 0.042 & 0.098\\
			& Cox \& Lee & 0.020 & 0.017 \\
			
			\hline
		\end{tabular}
	\end{table}
	
	\section{Application}
	In this section, we apply the funGP method to a wind energy problem. 
	A common technique to characterize the performance of a wind turbine is through the use of its power curve~\citep[Chapters 5 and 6]{Ding2019}. A univariate power curve is a functional curve with the wind speed as the input and the generated wind power as the output. But researchers realize that the wind power output is affected by other inputs more than just the wind speed.  Consequently, multivariate power curves have been developed; see, for instance, Chapter 5 of~\citet{Ding2019} or~\citet{Lee2015}.
	
\begin{figure}
		\centering
		\includegraphics[width = 3.5in]{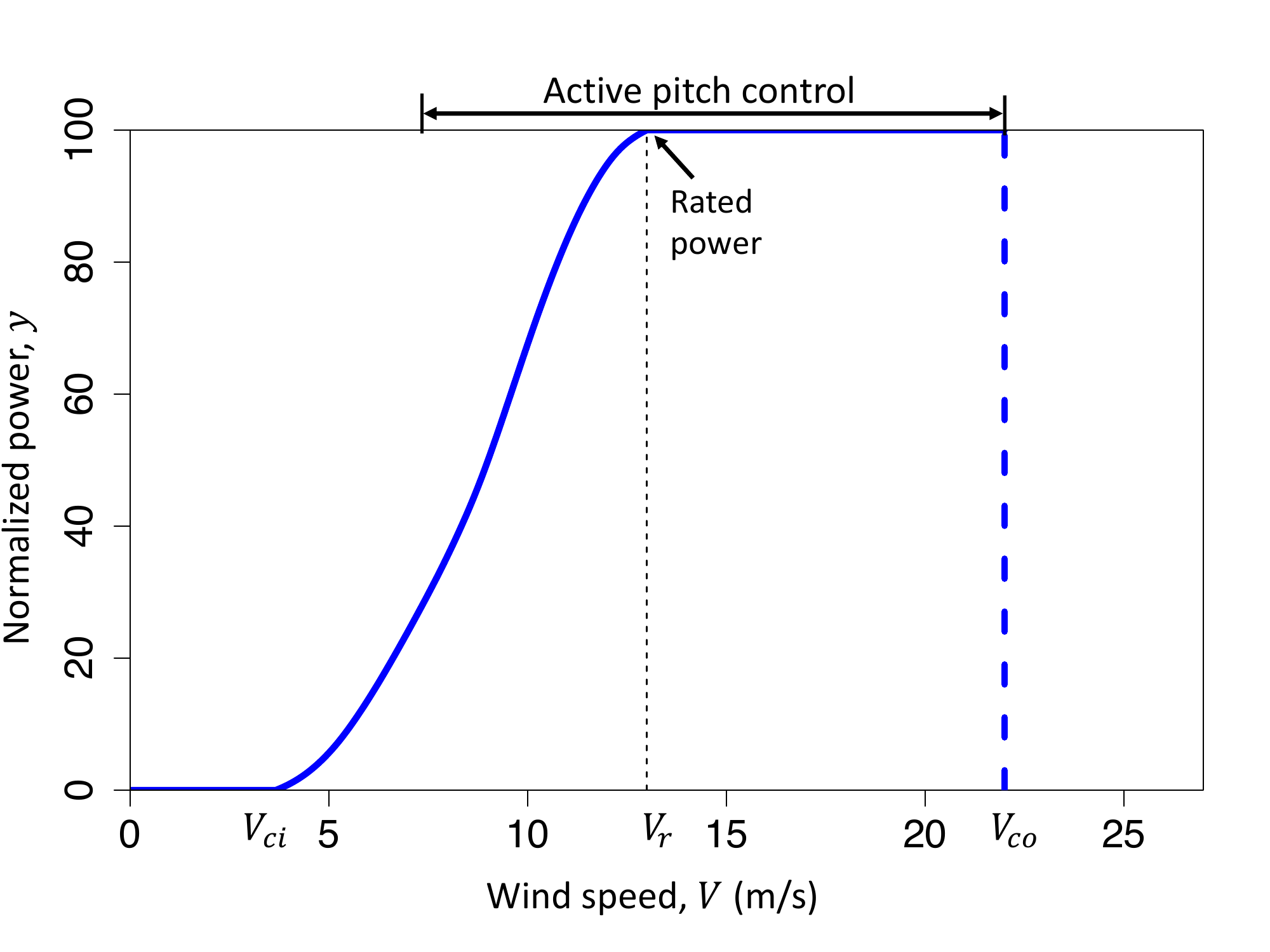}
		\caption{A nominal wind power curve. $V_{ci}$: the cut-in wind speed, $V_r$: the rated wind speed; $V_{co}$: the cut-out wind speed.}\label{fig:powercurve}
	\end{figure}

	A nominal wind power curve is shown in Figure~\ref{fig:powercurve}. The turbine does not produce power below the cut-in wind speed $V_{ci}$. Above the cut-in speed, the power gradually rises till the rated power and then capped at that level till the cut-out wind speed $V_{co}$, at which the turbine operation is stopped in order to protect its components against damage. The pitch control is one of the main mechanisms to regulate a wind turbine's power output~\citep{senjyu2006}; in Figure~\ref{fig:powercurve}, we mark the wind speed region where the pitch control is active.
	
	The power curve (univariate or multivariate) is generally learned through data; please see Chapter 5 of~\citet{Ding2019} for various methods. If one wants to compare the performance of  two turbines or the same turbine over multiple time periods, they can do so by comparing the learned power curves. This raises a question that whether the difference in the learned curves is due to the randomness in the samples, or the difference is genuine in turbine performance beyond random fluctuation. Our proposed method can, hence, be employed to answer this question. 
	
	We apply our method to the four datasets as used by~\cite{Hwangbo17}, which also constitutes a large portion of Chapter 6 of~\citet{Ding2019}, and we download the four datasets from the book website of~\citet{Ding2019}. Each dataset corresponds to a different turbine. The four turbines are labeled as WT1, WT2, WT3, and WT4. The datasets WT1 and WT2 are from onshore turbines and have the following five input variables: wind speed ($V$), wind direction ($ D $), air density ($ \rho $), turbulence intensity ($ I $), and wind shear ($ S $). The other two datasets (WT3 and WT4) correspond to offshore wind turbines with the input variable $ S $ replaced with humidity ($ H $), with the rest of the variables the same as that of the onshore turbines. Each of the four datasets comprises four years of data. We conduct a year to year comparison for each turbine, as done in~\cite{Hwangbo17}.  For this reason, each turbine's dataset is divided into four annual datasets.
	
	The marginal distributions of the covariates are different for each year, thus before computing their metric,~\cite{Hwangbo17} apply a method called covariate matching to the annual datasets. Covariate matching tries to match the marginal distributions of all the available environmental variables among the annual datasets by selecting the proper data subsets. Covariate matching is applied here in order to enable a fair comparison in turbine performance by ensuring that the distributions of the environmental variables are similar. We follow the same strategy with the same specifications as given in~\cite{Hwangbo17}. After the covariate matching,~\cite{Hwangbo17} uses only the wind speed as the input variable to estimate the power curve. We also proceed in a similar way. In other words, we have wind speed as the input and wind power as the output. We input these datasets to our funGP algorithm and do a pairwise comparison between the annual datasets for each turbine using the following specification. We select 1,000 evenly spaced points from the range of the input variable (wind speed) as the test grid and compare the power curves for any two annual datasets for a given turbine on the defined test grid. A typical wind turbine operates at wind speeds between 5 m/s to 15 m/s for most of the time. Thus, we select this range to test the difference.~\cite{Hwangbo17} developed a 90 \% confidence interval for their performance metric using the bootstrap method. For comparison, we also build a 90 \% confidence band on the difference of the power curves.
	
	The outputs from our method is the pointwise difference in the power curves and the 90 \% confidence band on the difference for the power curves to be the same. In Table~\ref{Table:percent_points}, we report the percentage of points, out of the 1,000 test points, where the difference between two given yearly datasets is statistically significant. Whenever the percentage is greater than zero, we claim that the difference between corresponding two curves is statistically significant.
	
	\begin{table}
		\spacingset{1}
		\caption{Percentage of test points with statistically significant difference between annual datasets.}\label{Table:percent_points}
		\vspace{6 pt}
		\centering
		\begin{tabular}{c| c c c c c c}
			\hline
			\textbf{Turbine} & Year 1 \& 2 & Year 1 \& 3  & Year 1 \& 4 & Year 2 \& 3 & Year 2 \& 4 & Year 3 \& 4\\
			\hline
			WT1 & 49.5 & 58.1 & 53.6 & 13.9 & 0 & 0 \\
			WT2 & 40.6 & 41.3 & 41.3 & 0 & 0 & 0 \\
			WT3 & 85.6 & 81.4  & 73.1 & 55.4 & 72.7 & 41.9 \\
			WT4  & 74.9 & 60.8 & 64.3 & 44.4 & 69.6 & 2.7 \\
			\hline			
		\end{tabular}
	\end{table}

	Speaking of the current industry practice for turbine performance comparison in the wind energy sector, the most popular method is to compare their peak power coefficient estimated from the data~\citep{IEC05}. The power coefficient, $C_p$, of a turbine is computed by using the following formula:
	\[C_p  = \frac{2y}{\rho A V^3},\]
	where $y$ is the wind power output and $A$ is the sweeping area of the turbine blades.  Here $C_p$ is not a constant but rather a function of wind speed and a few other factors.  The exact formula linking $C_p$ to other physical variables does not exist.  So it is empirically estimated.  Using a functional $C_p$ is not easy, and because of that, practitioners simply choose the peak value on the $C_p$-versus-wind-speed curve to represent the performance of a turbine. The power coefficient has a theoretical upper bound, known as the Betz limit, which is 0.593~\citep{Ding2019} but the practical $C_p$ is generally smaller than 0.5.  It is obvious that this $C_p$ metric is just a point metric of an otherwise functional difference.
	
	\cite{Hwangbo17} suggested another technique to compare the performance of wind turbines using the concepts of production economics. They devise a performance metric called productive efficiency which takes into account the overall power curve and not just the peak performance. But their  final output is again a point metric of the functional difference, much like the power coefficient. \cite{Hwangbo17}'s study find the productive efficiency metric has a good similarity with the power coefficient metric, although not exactly the same.  Using the four datasets mentioned above, the performance quantifications using the two metrics registered a correlation of 0.75~\citep{Hwangbo17}. Other than being a point metric, both the power coefficient and the productive efficiency methods do not quantify the estimation uncertainty on their own---one can go through an expensive bootstrap approach to get a confidence interval on the performance metrics. The funGP method, on the other hand, can lead to any level of confidence bands on the difference of the performance.
	
\begin{figure}[t]
		\centering
		\includegraphics[width=0.8\textwidth]{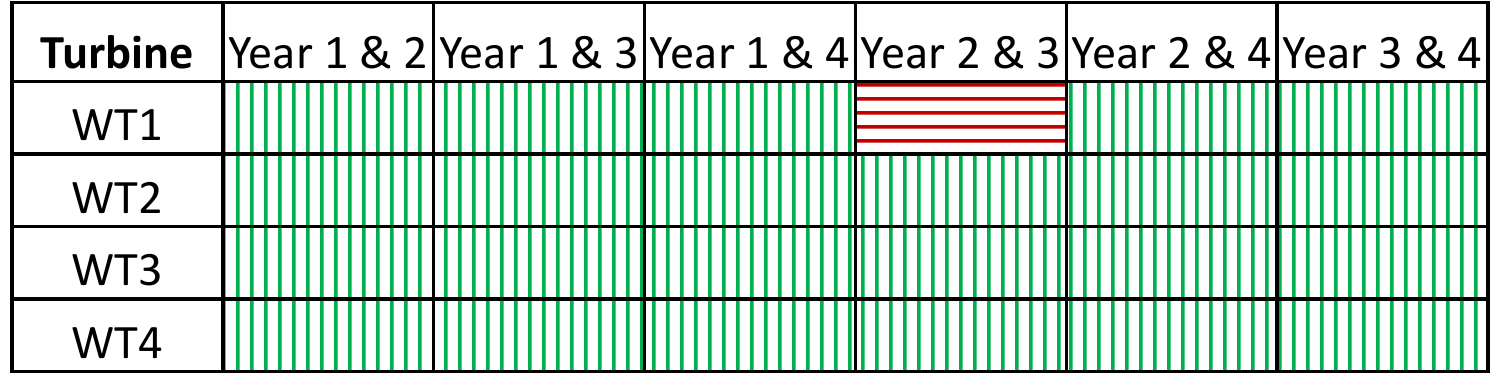}
		\caption{Comparison chart for the results obtained using funGP method to that of the peak power coefficient and the productive efficiency method. Vertical lines imply that the results agree. Horizontal lines imply that the results differ. } \label{ComparisonChart}
	\end{figure}

	We compare our results with the metrics, peak power coefficient and productive efficiency, obtained by~\citet[Table II]{Hwangbo17}. We illustrate the comparison in a chart (see Figure~\ref{ComparisonChart}) using vertical and horizontal lines with the following criteria:
	\begin{itemize}
		\spacingset{1}
		\item  If the two metrics used by~\citet{Hwangbo17} agree with each other (that is, they both say the two annual periods are different or they both say the same), and they also agree with our result, then we use vertical lines to demonstrate that.
		\item 	If the two metrics do not agree with each other, but one of them agree with our result, we still use vertical lines.
		\item However, if the two metrics agree with each other, but they do not agree with our method, we use horizontal lines to show that.
	\end{itemize}
	In other words, the vertical lines imply an agreement between our method and at least one of the two metrics, where as the horizontal lines mean a disagreement between the two metrics and the funGP method. We observe that when the difference between two power curves is statistically significant, the confidence intervals of the peak power coefficient or the productive efficiency for the same two curves tend not to overlap, leading naturally to the overwhelming agreement pattern observed in Figure~\ref{ComparisonChart}.
	
	There is one comparison outcome for which using funGP and either metric in~\citet{Hwangbo17} disagree: WT1 for Year 2 versus Year 3. Taking a closer look reveals that the percentages of test points where the two curves are different, as reported in Table~\ref{Table:percent_points}, is 13.9\%. The percentage is much smaller than the percentage values in other cases for which two curves are declared different. When we look at the power coefficient and productive efficiency values in~\citet[Table II]{Hwangbo17}, they are as such:
	\begin{itemize}
		\item WT 1's power coefficient.  Year 2: 0.388 with the 90\% confidence intervals as $[0.386, 0.392]$, and Year 3: 0.393 with the 90\% confidence intervals as $[0.390, 0.397]$.
		\item WT1's productive efficiency. Year 2: 0.969 with the 90\% confidence intervals as $[0.966, 0.973]$, and Year 3: 0.972 with the 90\% confidence intervals as $[0.969, 0.975]$.
	\end{itemize}
	Apparently, for the power coefficient and productive efficiency metrics, their 90\% confidence intervals are only marginally overlapping, not really contradicting with the small regions of difference detected by using the funGP method.  It is not unreasonable to consider that the funGP method is more sensitive to the difference between the two curves.
	
\begin{figure}
		\centering
		\includegraphics[width=2.5in]{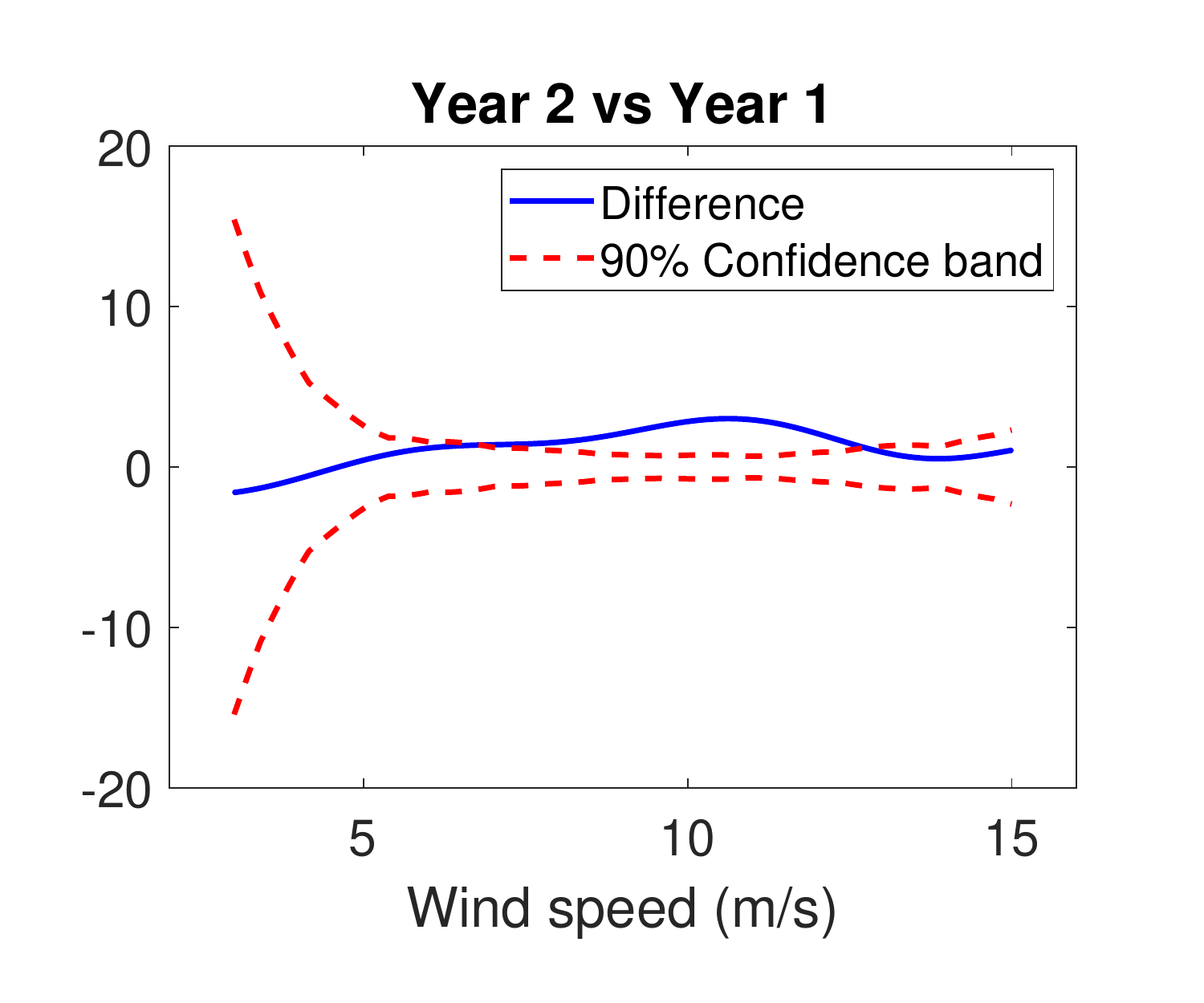}
		\includegraphics[width=2.5in]{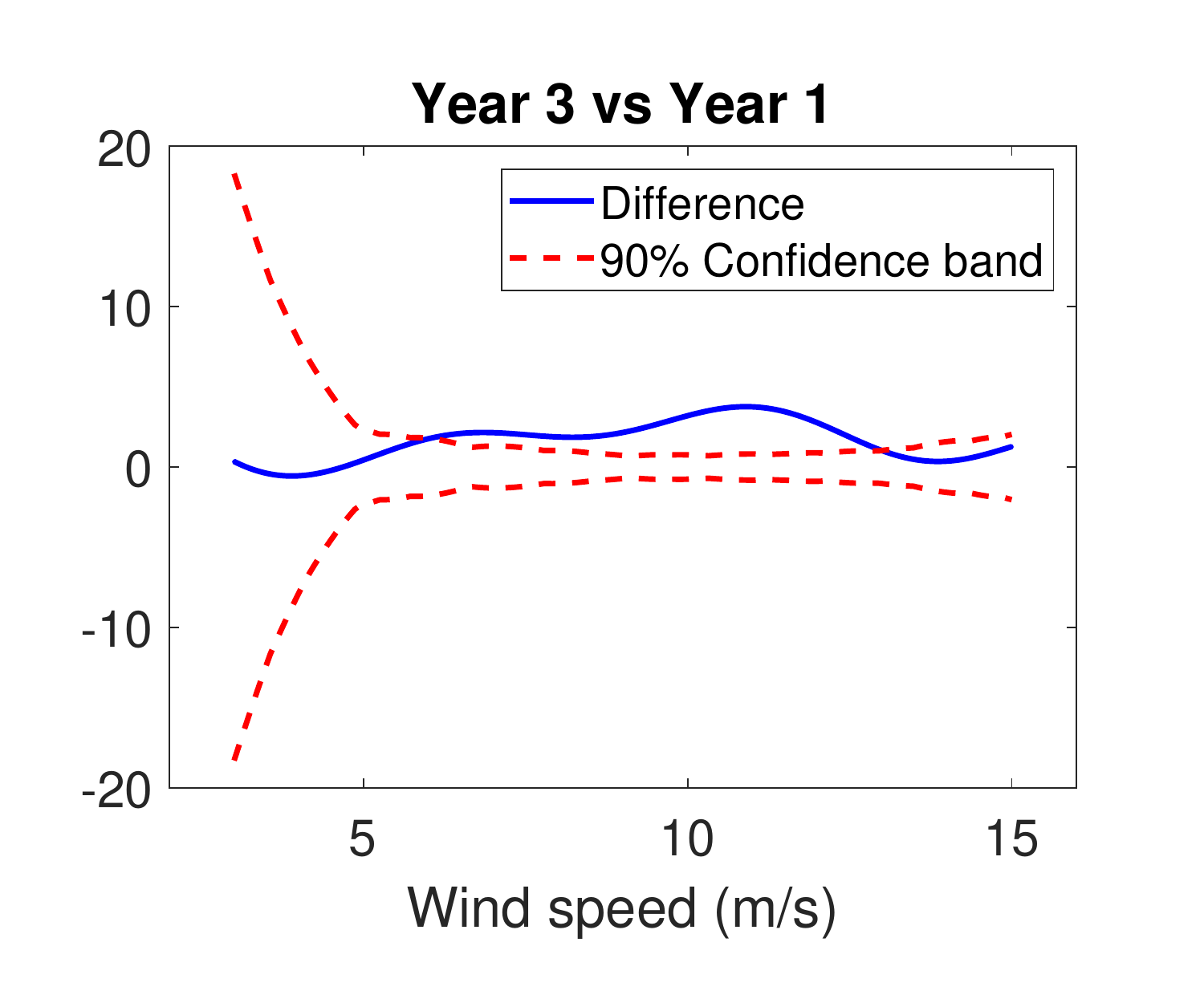}\\
		\includegraphics[width=2.5in]{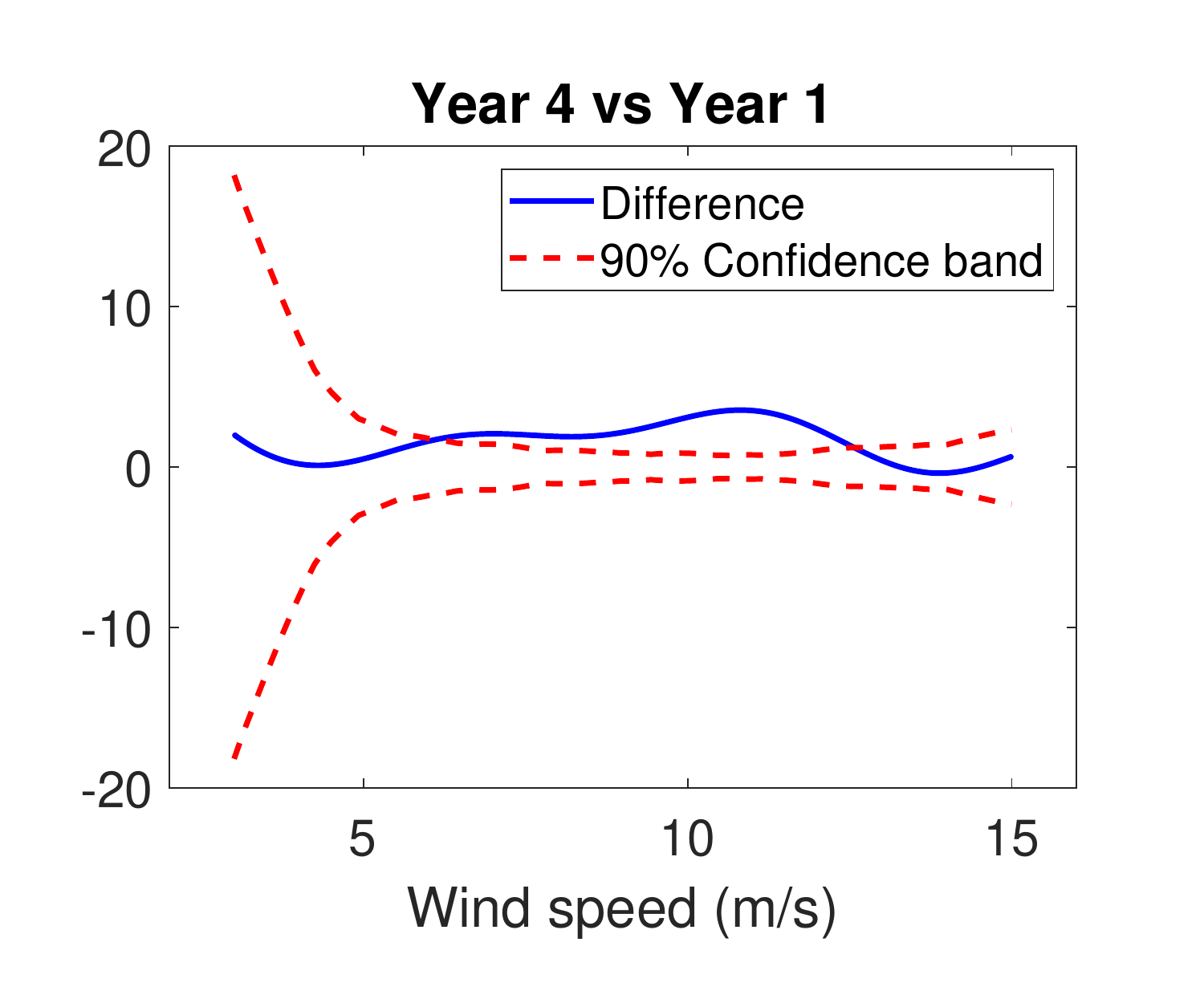}
		\includegraphics[width=2.5in]{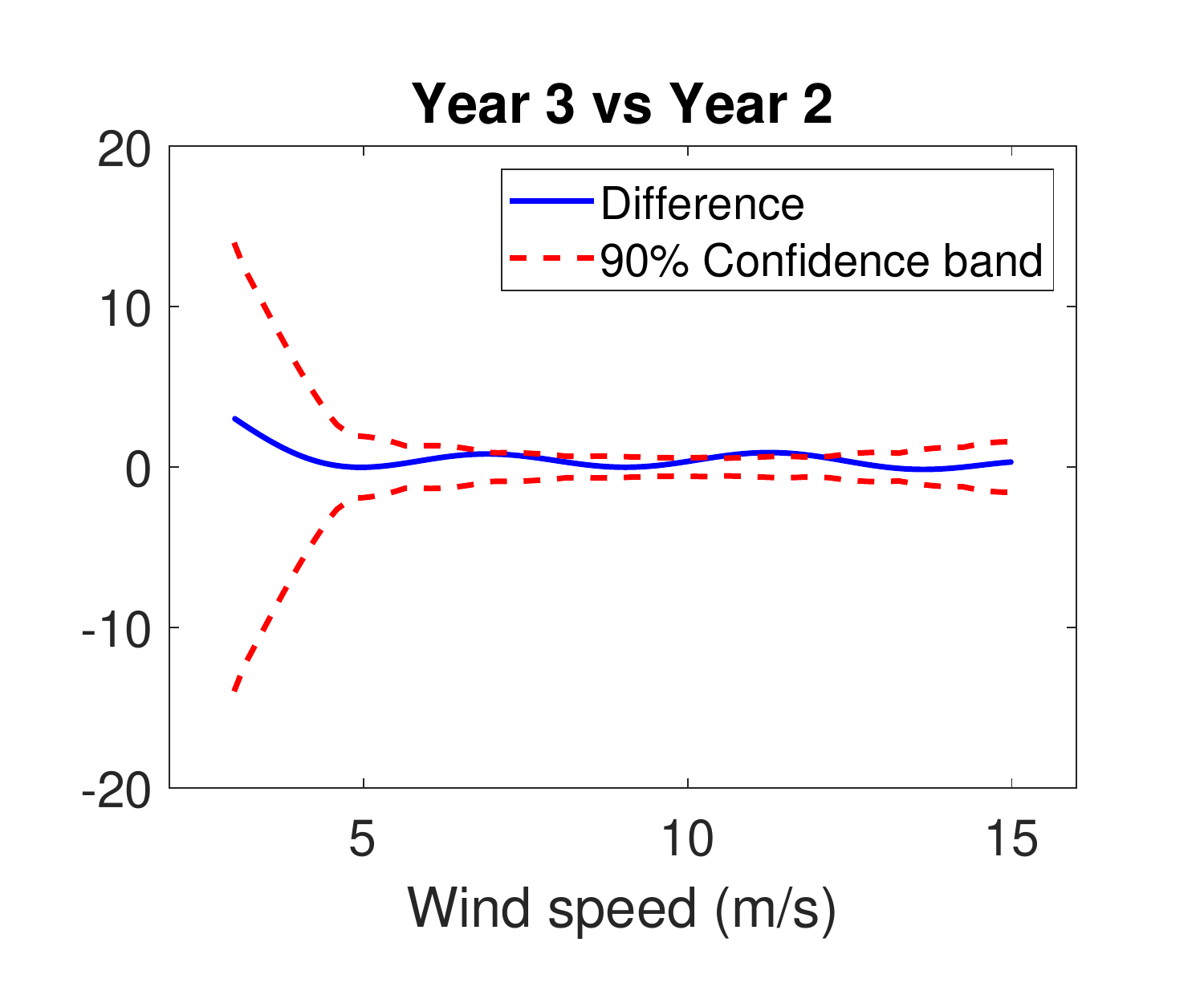}\\
		\includegraphics[width=2.5in]{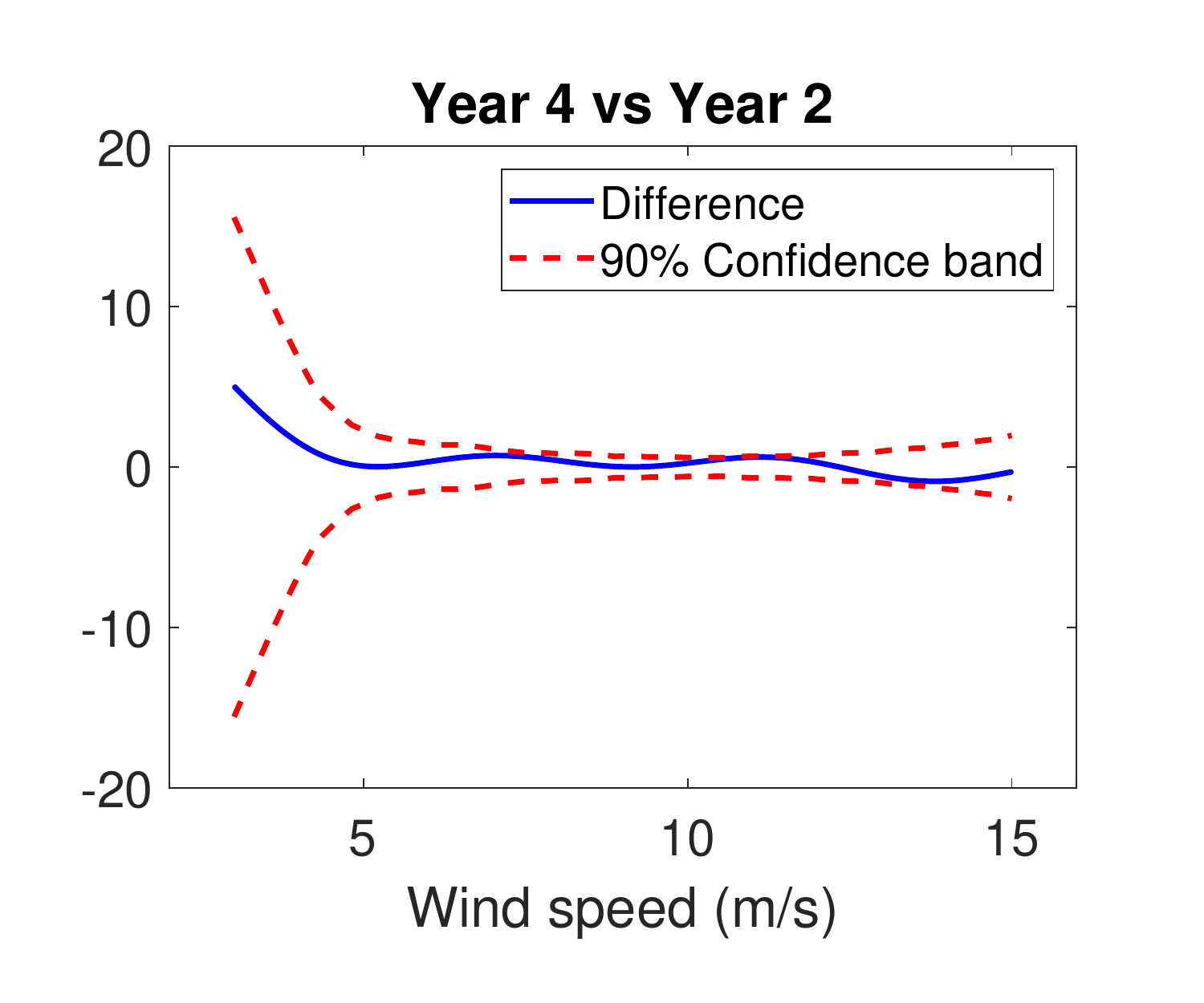}
		\includegraphics[width=2.5in]{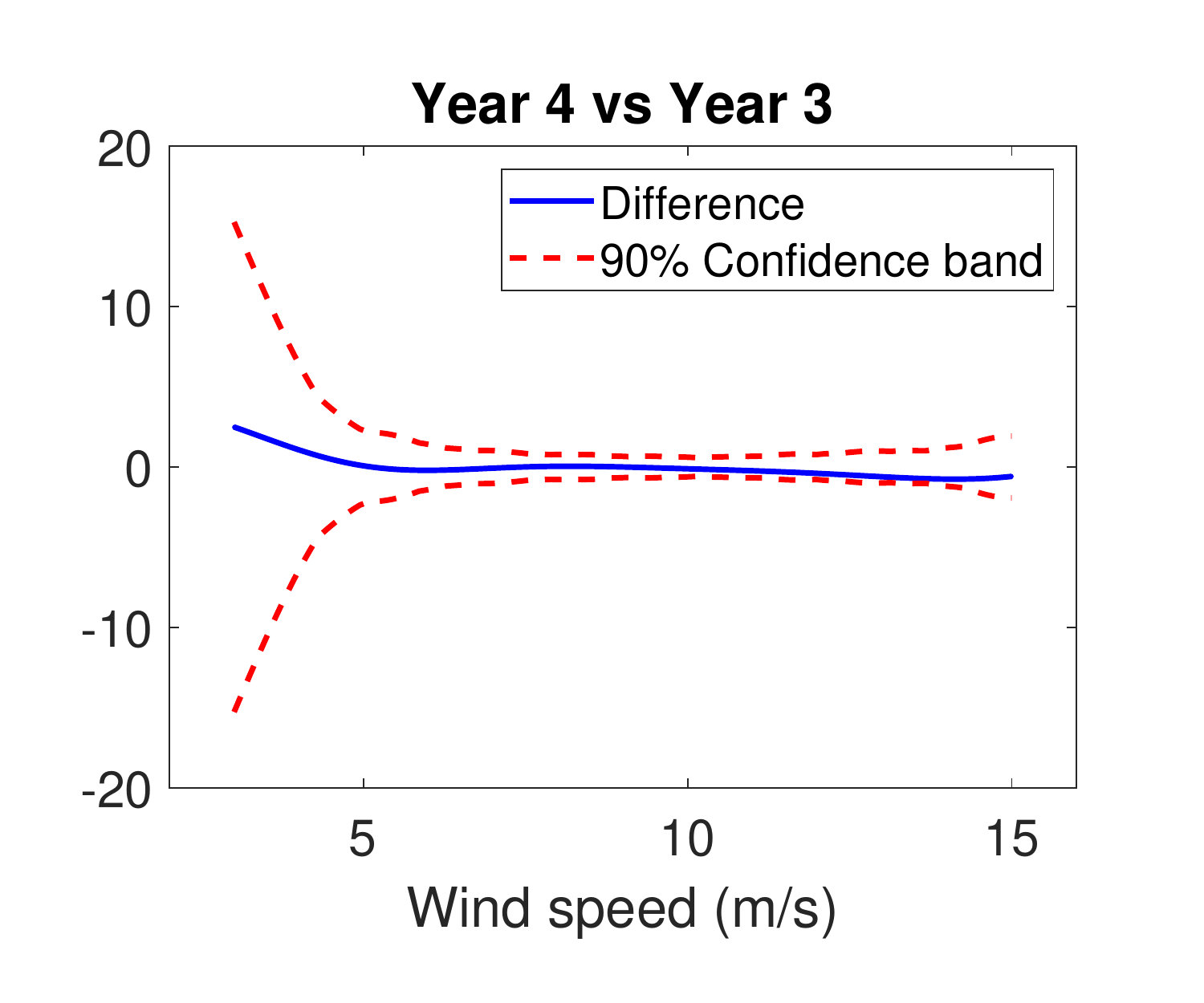}
		\caption{Difference in curves vs wind speed for WT1.}\label{Fig:DiffPlot}
	\end{figure}

	The funGP method provides a quantification of the regions of difference. Better yet, funGP can be used to compute the difference in the power curves at any point in the domain of the curve, and thus, gives a more detailed picture of the difference between any two curves, so that the practitioners can see where the difference lies and thus make an informed decision regarding whether the difference region matters or not. Figure~\ref{Fig:DiffPlot} shows this difference vs wind speed plot for all the annual datasets for the first turbine (WT1).
	
	As described in Section 1, knowing the regions of difference is helpful in deciding the maintenance plan for the turbine. For instance, if the difference occur in low power range, one may not necessarily need to go for expensive maintenance as doing so is unlikely to result in large change in the power output. Another important implication of knowing the difference regions is to decide the pitch control configuration of the turbine. As~\cite{Creaby2009} explains, wind turbine's aerodynamic characteristics change with time because of surface wear, dirt and other factors. Therefore, knowing the region of difference can help adjust the control laws to optimize the pitch control for different regions of operations, in order to maximize the power output. The funGP method is better suited in this application as a more powerful and informative testing and comparison method.
	
	\section{Discussions}
	This work presents a new nonparametric method that compare functions, referred to as the funGP method. Unlike many methods in the literature, the novelty of funGP lies in its ability to identify the regions of difference in the input space of the functions and quantify this difference, rather than simply returning a binary answer on whether the difference exists or not. This ability makes the funGP method a truly functional test.
	
	From an application point of view, particularly in engineering, comparing processes often mean subsequent decision making. For instance, comparing wind power curves guides the maintenance strategy. Under these circumstances, a binary answer for function comparison can easily run to its limit, as it may not be of much help in driving the decision making process. Understanding a fuller picture of function difference through region identification and subsequent quantification, on the other hand, could lead to better engineering and economic decisions. We hope that our work paves the way and shifts the focus of function comparison research towards more informative function tests, which would have broader applications and impact in the engineering fields.
	
	In the work, we use evenly spaced input points to conduct the comparison of the curves. In the higher dimensions, the numbers of grid points can grow rapidly and may become computationally burdensome. One worthy future research direction that would directly advance this work is to devise an adaptive grid---based on the characteristics of the function under study---to quickly identify and quantify the differences while reducing the computational time.

	
	%
	
	
	\if0\blind{
		\section*{Acknowledgment}
		Prakash and Ding's research is partially supported by NSF grant IIS-1741173. Tuo's research is supported by NSF DMS-1914636. Ding and Tuo's research is also supported by NSF grant CCF-1934904.
	} \fi
	
	\baselineskip=15 pt
	\bibliographystyle{apalike}
	\bibliography{TwoFunctionTest}

\begin{thebibliography}{}

\bibitem[Cox and Lee, 2008]{CoxLee2008}
Cox, D. and Lee, J.~S. (2008).
\newblock Pointwise testing with functional data using the {W}estfall--{Y}oung
  randomization method.
\newblock {\em Biometrika}, 95(3):621--634.

\bibitem[Creaby et~al., 2009]{Creaby2009}
Creaby, J., Li, Y., and Seem, J.~E. (2009).
\newblock Maximizing wind turbine energy capture using multivariable extremum
  seeking control.
\newblock {\em Wind Engineering}, 33(4):361--387.

\bibitem[Delgado, 1993]{Delgado1993}
Delgado, M.~A. (1993).
\newblock Testing the equality of nonparametric regression curves.
\newblock {\em Statistics \& Probability Letters}, 17(3):199--204.

\bibitem[Ding, 2019]{Ding2019}
Ding, Y. (2019).
\newblock {\em Data Science for Wind Energy}.
\newblock Chapman \& Hall/CRC Press, Boca Raton, FL.

\bibitem[Fan and Lin, 1998]{FanLin98}
Fan, J. and Lin, S.-K. (1998).
\newblock Test of significance when data are curves.
\newblock {\em Journal of the American Statistical Association},
  93(443):1007--1021.

\bibitem[Fisher, 1925]{fisher1925}
Fisher, R.~A. (1925).
\newblock Application of ``{S}tudent's" distribution.
\newblock {\em Metron}, 5:90--104.

\bibitem[Hall and Hart, 1990]{HallHart1990}
Hall, P. and Hart, J.~D. (1990).
\newblock Bootstrap test for difference between means in nonparametric
  regression.
\newblock {\em Journal of the American Statistical Association},
  85(412):1039--1049.

\bibitem[Harper and Gupta, 1983]{harper1983}
Harper, W. and Gupta, S. (1983).
\newblock {\em Sensitivity/uncertainty analysis of a borehole scenario
  comparing {L}atin Hypercube Sampling and deterministic sensitivity
  approaches}.
\newblock BMI/ONWI-516, Office of Nuclear Waste Isolation, Battelle Memorial
  Institute, Columbus, OH.

\bibitem[Hotelling, 1931]{hotelling1931}
Hotelling, H. (1931).
\newblock The generalization of {S}tudent's ratio.
\newblock {\em The Annals of Mathematical Statistics}, 2(3):360--378.

\bibitem[Hwangbo et~al., 2017]{Hwangbo17}
Hwangbo, H., Johnson, A., and Ding, Y. (2017).
\newblock A production economics analysis for quantifying the efficiency of
  wind turbines.
\newblock {\em Wind Energy}, 20(9):1501--1513.

\bibitem[IEC, 2005]{IEC05}
IEC (2005).
\newblock {\em Wind Turbines-Part 12-1: Power Performance Measurements of
  Electricity Producing Wind Turbines}.
\newblock International Electrotechnical Commission 61400-12-1 Ed. 1, Geneva,
  Switzerland.

\bibitem[Kenett and Zacks, 1998]{Kenett_Zacks_1998}
Kenett, R.~S. and Zacks, S. (1998).
\newblock {\em Modern Industrial Statistics: The Design and Control of Quality
  and Reliability}.
\newblock Duxbury Press, Pacific Grove, CA.

\bibitem[King et~al., 1991]{King1991}
King, E., Hart, J.~D., and Wehrly, T.~E. (1991).
\newblock Testing the equality of two regression curves using linear smoothers.
\newblock {\em Statistics \& Probability Letters}, 12(3):239--247.

\bibitem[Kulasekera, 1995]{kulasekara1995}
Kulasekera, K.~B. (1995).
\newblock Comparison of regression curves using quasi-residuals.
\newblock {\em Journal of the American Statistical Association},
  90(431):1085--1093.

\bibitem[Kulasekera and Wang, 1997]{kulasekara1997}
Kulasekera, K.~B. and Wang, J. (1997).
\newblock Smoothing parameter selection for power optimality in testing of
  regression curves.
\newblock {\em Journal of the American Statistical Association},
  92(438):500--511.

\bibitem[Lee et~al., 2015]{Lee2015}
Lee, G., Ding, Y., Genton, M.~G., and Xie, L. (2015).
\newblock Power curve estimation with multivariate environmental factors for
  inland and offshore wind farms.
\newblock {\em Journal of the American Statistical Association},
  110(509):56--67.

\bibitem[Morris et~al., 1993]{Morris-et-al_1993}
Morris, M.~D., Mitchell, T.~J., and Ylvisaker, D. (1993).
\newblock Bayesian design and analysis of computer experiments: Use of
  derivatives in surface prediction.
\newblock {\em Technometrics}, 35(3):243--255.

\bibitem[Munk and Dette, 1998]{munk1998}
Munk, A. and Dette, H. (1998).
\newblock Nonparametric comparison of several regression functions: {E}xact and
  asymptotic theory.
\newblock {\em The Annals of Statistics}, 26(6):2339--2368.

\bibitem[Neumeyer and Dette, 2003]{neumeyer2003}
Neumeyer, N. and Dette, H. (2003).
\newblock Nonparametric comparison of regression curves: {A}n empirical process
  approach.
\newblock {\em The Annals of Statistics}, 31(3):880--920.

\bibitem[Rasmussen and Williams, 2006]{Rasmussen2006}
Rasmussen, C.~E. and Williams, C. K.~I. (2006).
\newblock {\em Gaussian Processes for Machine Learning}.
\newblock The MIT Press, Cambridge, MA.

\bibitem[Senjyu et~al., 2006]{senjyu2006}
Senjyu, T., Sakamoto, R., Urasaki, N., Funabashi, T., Fujita, H., and Sekine,
  H. (2006).
\newblock Output power leveling of wind turbine generator for all operating
  regions by pitch angle control.
\newblock {\em IEEE Transactions on Energy conversion}, 21(2):467--475.

\bibitem[Westfall and Young, 1993]{westfall1993}
Westfall, P.~H. and Young, S.~S. (1993).
\newblock {\em Resampling-Based Multiple Testing: Examples and Methods for
  p-Value Adjustment}.
\newblock John Wiley \& Sons, New York, NY.

\end{thebibliography}
	\section*{Appendix}
	\spacingset{2}
	\subsection*{A.1 Derivation for $c(\bs{x},\bs{x}')$}
	The predictive mean for $ f_1(.) $ given $\mathcal{D}_1$ is as follows:
	\begin{equation*}
		\hat{f}_1(\boldsymbol{x}) = \boldsymbol{r}_1(\boldsymbol{x})^\top[\mathbf{K}_{\mb{X}^{(1)},\mb{X}^{(1)}}+\sigma_{\epsilon}^2\mathbf{I}_{n_1}]^{-1}\boldsymbol{y}^{(1)}.
	\end{equation*}
	Similarly, the predictive mean for $ f_2(.) $ conditioned on $\mathcal{D}_2$ is given by:
	\begin{equation*}
		\hat{f}_2(\boldsymbol{x}) = \boldsymbol{r}_2(\boldsymbol{x})^\top[\mathbf{K}_{\mb{X}^{(2)},\mb{X}^{(2)}}+\sigma_{\epsilon}^2\mathbf{I}_{n_2}]^{-1}\boldsymbol{y}^{(2)}.
	\end{equation*}
	Thus $ c(\bs{x},\bs{x}') = Cov ( \hat{f}_2(\bs{x}) - \hat{f}_1(\bs{x}) )$ is expressed as follows:
	\begin{equation*}
		\begin{split}
			&Cov ( \hat{f}_2(\bs{x}) - \hat{f}_1(\bs{x}) )\\
			& = Cov(\boldsymbol{r}_2(\boldsymbol{x})^\top[\mathbf{K}_{\mb{X}^{(2)},\mb{X}^{(2)}}+\sigma_{\epsilon}^2\mathbf{I}_{n_2}]^{-1}\boldsymbol{y}^{(2)} - \boldsymbol{r}_1(\boldsymbol{x})^\top[\mathbf{K}_{\mb{X}^{(1)},\mb{X}^{(1)}}+\sigma_{\epsilon}^2\mathbf{I}_{n_1}]^{-1}\boldsymbol{y}^{(1)})\\
			\\
			& = Var( \boldsymbol{r}_2(\boldsymbol{x})^\top[\mathbf{K}_{\mb{X}^{(2)},\mb{X}^{(2)}}+\sigma_{\epsilon}^2\mathbf{I}_{n_2}]^{-1}\boldsymbol{y}^{(2)})
			+ Var(\boldsymbol{r}_1(\boldsymbol{x})^\top[\mathbf{K}_{\mb{X}^{(1)},\mb{X}^{(1)}}+\sigma_{\epsilon}^2\mathbf{I}_{n_1}]^{-1}\boldsymbol{y}^{(1)}) \\
			& - 2 \ Cov( \boldsymbol{r}_2(\boldsymbol{x})^\top[\mathbf{K}_{\mb{X}^{(2)},\mb{X}^{(2)}}+\sigma_{\epsilon}^2\mathbf{I}_{n_2}]^{-1}\boldsymbol{y}^{(2)},\boldsymbol{r}_1(\boldsymbol{x})^\top[\mathbf{K}_{\mb{X}^{(1)},\mb{X}^{(1)}}+\sigma_{\epsilon}^2\mathbf{I}_{n_1}]^{-1}\boldsymbol{y}^{(1)} ) \\
			\\
			& =  \boldsymbol{r}_2(\boldsymbol{x})^\top[\mathbf{K}_{\mb{X}^{(2)},\mb{X}^{(2)}}+\sigma_{\epsilon}^2\mathbf{I}_{n_2}]^{-1}  \ Var(\boldsymbol{y}^{(2)}) \ [\mathbf{K}_{\mb{X}^{(2)},\mb{X}^{(2)}}+\sigma_{\epsilon}^2\mathbf{I}_{n_2}]^{-1}\boldsymbol{r}_2(\boldsymbol{x}')\\
			& +\boldsymbol{r}_1(\boldsymbol{x})^\top[\mathbf{K}_{\mb{X}^{(1)},\mb{X}^{(1)}}+\sigma_{\epsilon}^2\mathbf{I}_{n_1}]^{-1}  \ Var(\boldsymbol{y}^{(1)}) \ [\mathbf{K}_{\mb{X}^{(1)},\mb{X}^{(1)}}+\sigma_{\epsilon}^2\mathbf{I}_{n_1}]^{-1}\boldsymbol{r}_1(\boldsymbol{x}')\\
			& - 2 \  \boldsymbol{r}_2(\boldsymbol{x})^\top[\mathbf{K}_{\mb{X}^{(2)},\mb{X}^{(2)}}+\sigma_{\epsilon}^2\mathbf{I}_{n_2}]^{-1}  \ Cov(\boldsymbol{y}^{(2)},\boldsymbol{y}^{(1)}) \ [\mathbf{K}_{\mb{X}^{(1)},\mb{X}^{(1)}}+\sigma_{\epsilon}^2\mathbf{I}_{n_1}]^{-1}\boldsymbol{r}_1(\boldsymbol{x}')\\
			\\
			& =   \boldsymbol{r}_2(\boldsymbol{x})^\top[\mathbf{K}_{\mb{X}^{(2)},\mb{X}^{(2)}}+\sigma_{\epsilon}^2\mathbf{I}_{n_2}]^{-1}\ \boldsymbol{r}_2(\boldsymbol{x}')
			+\boldsymbol{r}_1(\boldsymbol{x})^\top \ [\mathbf{K}_{\mb{X}^{(1)},\mb{X}^{(1)}}+\sigma_{\epsilon}^2\mathbf{I}_{n_1}]^{-1}\boldsymbol{r}_1(\boldsymbol{x}')\\
			& - 2 \  \boldsymbol{r}_2(\boldsymbol{x})^\top[\mathbf{K}_{\mb{X}^{(2)},\mb{X}^{(2)}}+\sigma_{\epsilon}^2\mathbf{I}_{n_2}]^{-1}  \ \mb{K}_{\mb{X}^{(2)},\mb{X}^{(1)}} \ [\mathbf{K}_{\mb{X}^{(1)},\mb{X}^{(1)}}+\sigma_{\epsilon}^2\mathbf{I}_{n_1}]^{-1}\boldsymbol{r}_1(\boldsymbol{x}').
		\end{split}
	\end{equation*}
	
	\subsection*{A.2 Karhunen-Lo\`{e}ve expansion of a Gaussian process}
	Karhunen-Lo\`{e}ve expansion provides a framework to decompose any stochastic process as an infinite linear combination of orthogonal basis functions. Since, we are interested in Gaussian processes, we will discuss the KL expansion only for GPs. Let us now consider that $f(\boldsymbol{x})$ is a zero mean Gaussian process with $k(\boldsymbol{x},\boldsymbol{x}')$ as the covariance function. This process can decomposed as follows:
	\begin{equation}
		f(\boldsymbol{x}) = \sum_{k=1}^{\infty}\sqrt{\lambda_k} \phi_k(\boldsymbol{x}) z_k,
	\end{equation}
	where $z_k\ |\ k = 1,\dots,\infty$ are the uncorrelated standard normal random variables, $\lambda_k \ | \ k = 1,\dots,\infty$ are the eigenvalues,
	and $\phi_k(.)\ |\ k = 1,\dots,\infty$ are the basis eigenfunctions.
	The values of $\lambda_k$ and $\phi_k(.)$ can be obtained by solving the following integral eigenproblem
	\begin{equation}\label{IntEigenProb}
		\int k(\boldsymbol{x},\boldsymbol{x}')\phi(\boldsymbol{x}')d\boldsymbol{x}' = \lambda\phi(\boldsymbol{x}).
	\end{equation}
	In practice, Equation \eqref{IntEigenProb} can be solved by discretizing the integral. Let us again assume that we have $ n $ data points from the process $ f(\cdot) $. Then, we consider the following matrix eigenproblem
	\begin{equation}
		\mathbf{K}\boldsymbol{u}_k = \lambda_k^{mat} \boldsymbol{u}_k,
	\end{equation}
	where $\mathbf{K}$ is again the covariance matrix with entries $\mathbf{K}_{ij} = k(\boldsymbol{x}_i,\boldsymbol{x}_j) \ | \ i,j = 1 \dots n $;\\
	$\lambda_k^{mat}$ are the eigenvalues of the covariance matrix $\mathbf{K}$;\\
	$\boldsymbol{u}_k$ are the normalized unit eigenvectors of the covariance matrix $\mathbf{K}$.\\
	The eigenvalues and eigenfunctions of the integral problem are related to the eigenvalues and eigenvectors of the matrix problem in the following way:
	\begin{eqnarray}
		\lambda_k & \approx & \frac{\lambda_k^{mat}}{n},\\
		\phi_k(\boldsymbol{x}_j) & \approx & \sqrt{n}(\boldsymbol{u}_k)_j	,
	\end{eqnarray}
	where $ (\boldsymbol{u}_k)_j	 $ is the $ j^{th} $ component of the eigenvector $ \boldsymbol{u}_k $.
	The above approximation reduces the infinite sum in the KL expansion to a finite sum (truncated KL expansion) as follows:
	\begin{equation}
		\begin{split}
			f(x_j) & \approx \sum_{k=1}^{n}\sqrt{\frac{\lambda_k^{mat}}{n}} \sqrt{n}(\boldsymbol{u}_k)_j z_k, \\
			& = \sum_{k=1}^{n}\sqrt{\lambda_k^{mat}} (\boldsymbol{u}_k)_j z_k.
		\end{split}
	\end{equation}
	If $lambda_k$'s decay rapidly, this sum can be be truncated further by considering only $m$ largest eigenvalues, where $m < n$. This decomposition can be written compactly in the matrix form. If we consider a vector, $ \boldsymbol{f} = (f(\boldsymbol{x}_1),f(\boldsymbol{x}_2),\dots, f(\boldsymbol{x}_n)) ^\top$, then it can be decomposed as follows:
	\begin{equation}
		\boldsymbol{f} = \mathbf{U}\mathbf{\Lambda}^{\frac{1}{2}}\boldsymbol{z},
	\end{equation}
	where $\mathbf{U}$ is the matrix with columns as eigenvectors of covariance matrix $ \mathbf{K} $; $\mathbf{\Lambda}$ is a diagonal matrix with eigenvalues of $ \mathbf{K} $ and $\boldsymbol{z}$ is a vector of length $n$ with uncorrelated standard normal random variables as its components.
	
	\subsection*{A.3 funGP algorithm}
	
	\begin{algorithm}[H]
		\spacingset{1}
		\caption{funGP: function comparison using Gaussian process}
		\SetAlgoLined
		\textbf{Input:} {$\mathcal{D}_1 = \{\mathbf{X}^{(1)},\boldsymbol{y}^{(1)} \} $, $\mathcal{D}_2=\{\mathbf{X}^{(2)},\boldsymbol{y}^{(2)}\}$, $\mathbf{X}_{test}$}, $ \alpha $\\
		\textbf{Procedure:}\\
		1: Choose a covariance function.\\
		2: Estimate the hyperparameters for the covariance function and the nugget, $ \sigma_{\epsilon} $, by optimizing the likelihood function given in Equation \eqref{Eqn:Likelihood}.\\
		3: Compute the predictive mean functions $ \hat{f}_1 $ using $ \mathcal{D}_1 $, and $\hat{f}_2 $ using $ \mathcal{D}_2$ using Equations \eqref{posteroirFn1} and \eqref{posteroirFn2}.\\
		4: Compute the covariance matrix $ \mathbf{C}_{\mb{X}_{test},\mb{X}_{test}} $ using the covariance function in Equation \eqref{DiffCov} for the points in $\mb{X}_{test}$.\\
		4: Compute the difference between predictive means for the points in $\mb{X}_{test}$, $ g(\bs{x}_{t_j}) = \hat{f}_2(\bs{x}_{t_j}) - \hat{f}_1(\bs{x}_{t_j}) \ | \ j = 1,\dots,n_{test} $.\\
		6: Do the eigen decomposition of $ \mathbf{C}_{\mb{X}_{test},\mb{X}_{test}} $ and store the $m$ largest eigenvalues following the truncation rule in Section \ref{Sec2.3} in a diagonal matrix $ {\mathbf{\Lambda}} $ and the corresponding eigenvectors in a matrix  $ {\mathbf{U}} $.\\
		7: Compute the radius, $ r $, of a standard normal vector of dimension $ m $ with a coverage probability of $ 1-\alpha $ using Equation \eqref{Eqn:radius}. \\
		8: Sample a large number (say 1,000) of standard normal vector $ \boldsymbol{z} $ such that $ ||\boldsymbol{z}|| \leq r$.\\
		9: Compute the vector of upper bounds, $\boldsymbol{ub}$, and lower bounds, $\boldsymbol{lb}$, for all the test points using Equation \eqref{Eqn:band}.\\
		\textbf{Output:}\\
		If $(\boldsymbol{lb})_j \leq g(\bs{x}_{t_j}) \leq (\boldsymbol{ub})_j \quad \forall \ j = 1,\dots, n_{test}$, \emph{functions are same at $ 1 -\alpha $ confidence level.}\\
		Else, \emph{functions are different at $ 1 -\alpha $ confidence level.}
	\end{algorithm}
	
	\subsection*{A.4 Details of the simulated functions}
	\textbf{Piston simulation function}\\
	The piston simulation function, as the name suggests, is used to simulate the motion of a piston inside an engine.  This function was proposed by~\cite{Kenett_Zacks_1998}. The response is the cycle time in seconds, i.e. the time required to complete one cycle, and is given by:
	\[
	f(x) = 2\pi \sqrt{\frac{M}{k+S^2\frac{P_0V_0}{T_0}\frac{T_a}{V^2}}},
	\]
	where
	\[
	V = \frac{S}{2k}\Bigg(\sqrt{A^2+4k\frac{P_0V_0}{T_0}T_a}-A\Bigg),
	\]
	\[
	A = P_0S + 19.62M -\frac{kV_0}{S},
	\]
	where $ M $ is the weight of the piston  ($ kg $), $ k $ is the coefficient of the spring,
	$ S $ is the piston surface area ($ m^2 $) , $ P_0 $ is the atmospheric pressure ($N/m^2$),
	$ V_0 $ is the initial gas volume ($ m^3 $), $ T_0 $ is the filling gas temperature ($ K $),
	and $ T_a $ is the ambient temperature ($ K $).
	The number of input variables in this function are seven. We only choose two of them, $ V_0 $ and $ T_0 $, as input variables. The other variables are fixed at $M = 45$, $S = 0.01$, $k = 2,000$, $P_0 = 100,000$, $T_a = 292$.
	\begin{figure}
		\centering
		\includegraphics[width = 2.5in]{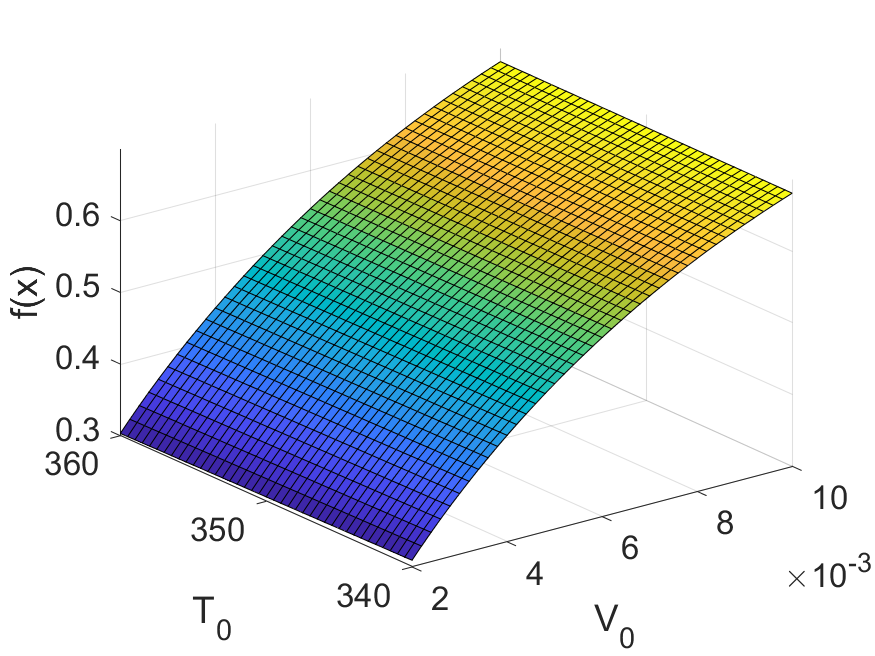}\qquad
		\includegraphics[width = 2.5in]{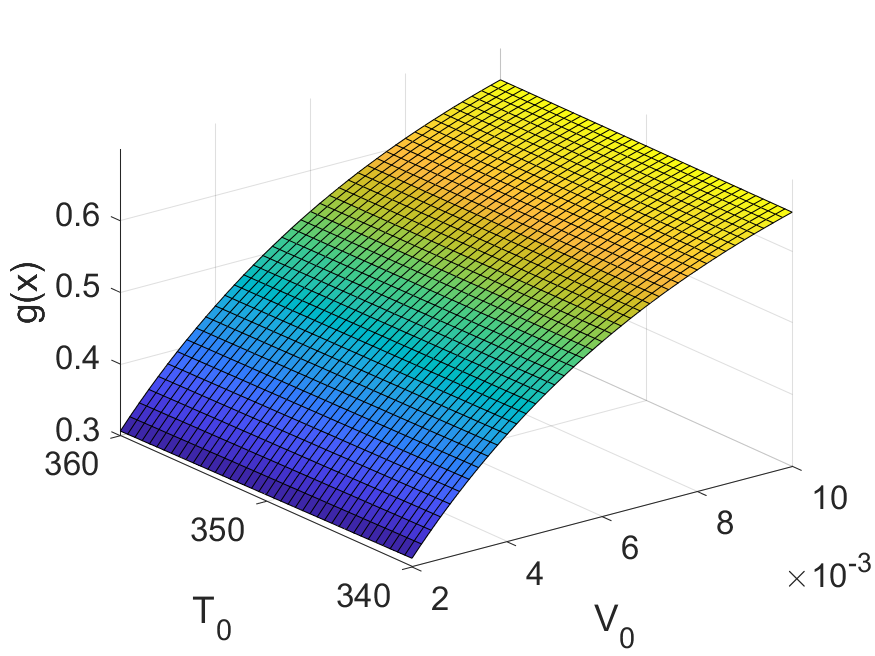}\\
		\includegraphics[width = 2.75in]{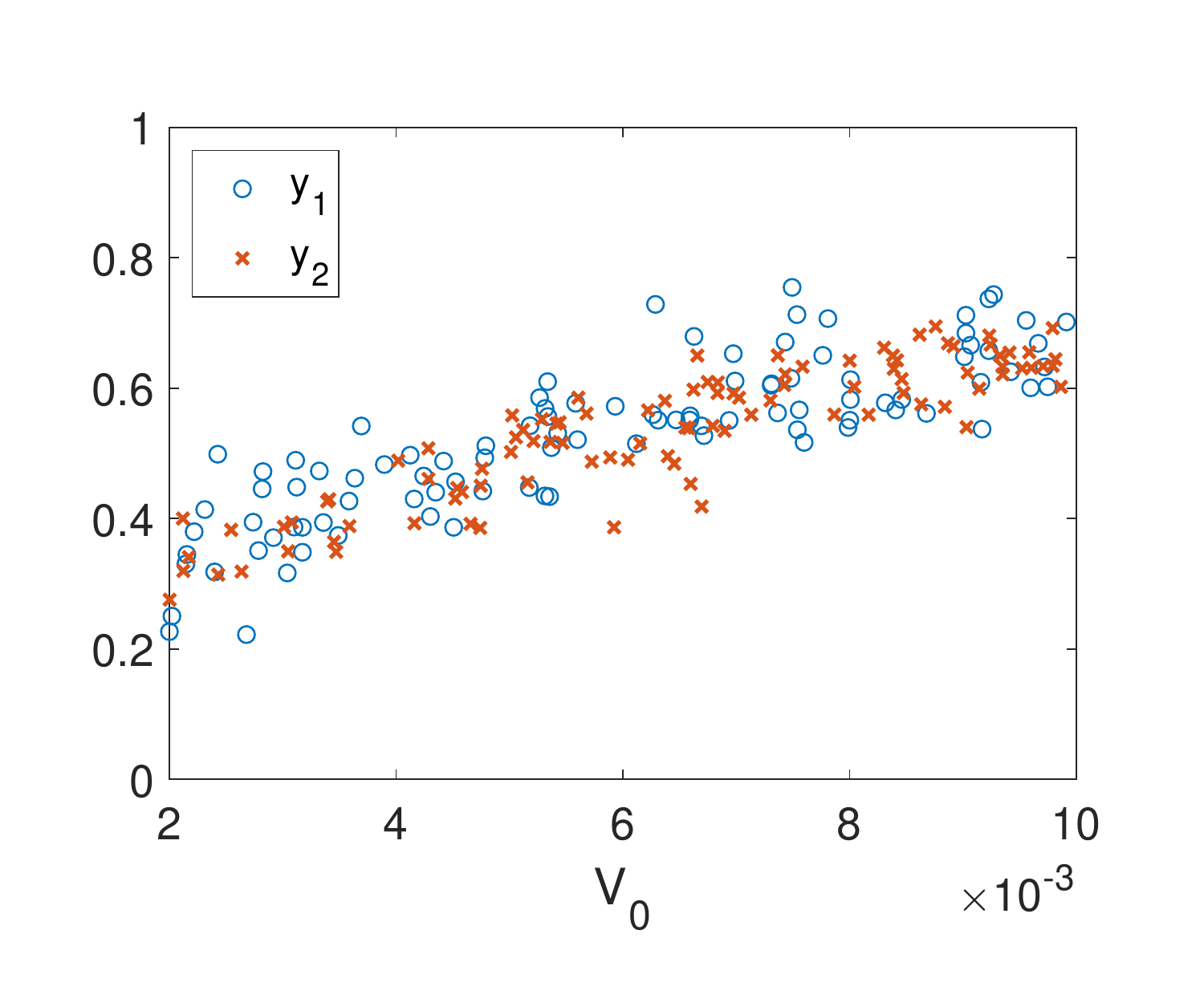}\qquad
		\includegraphics[width = 2.75in]{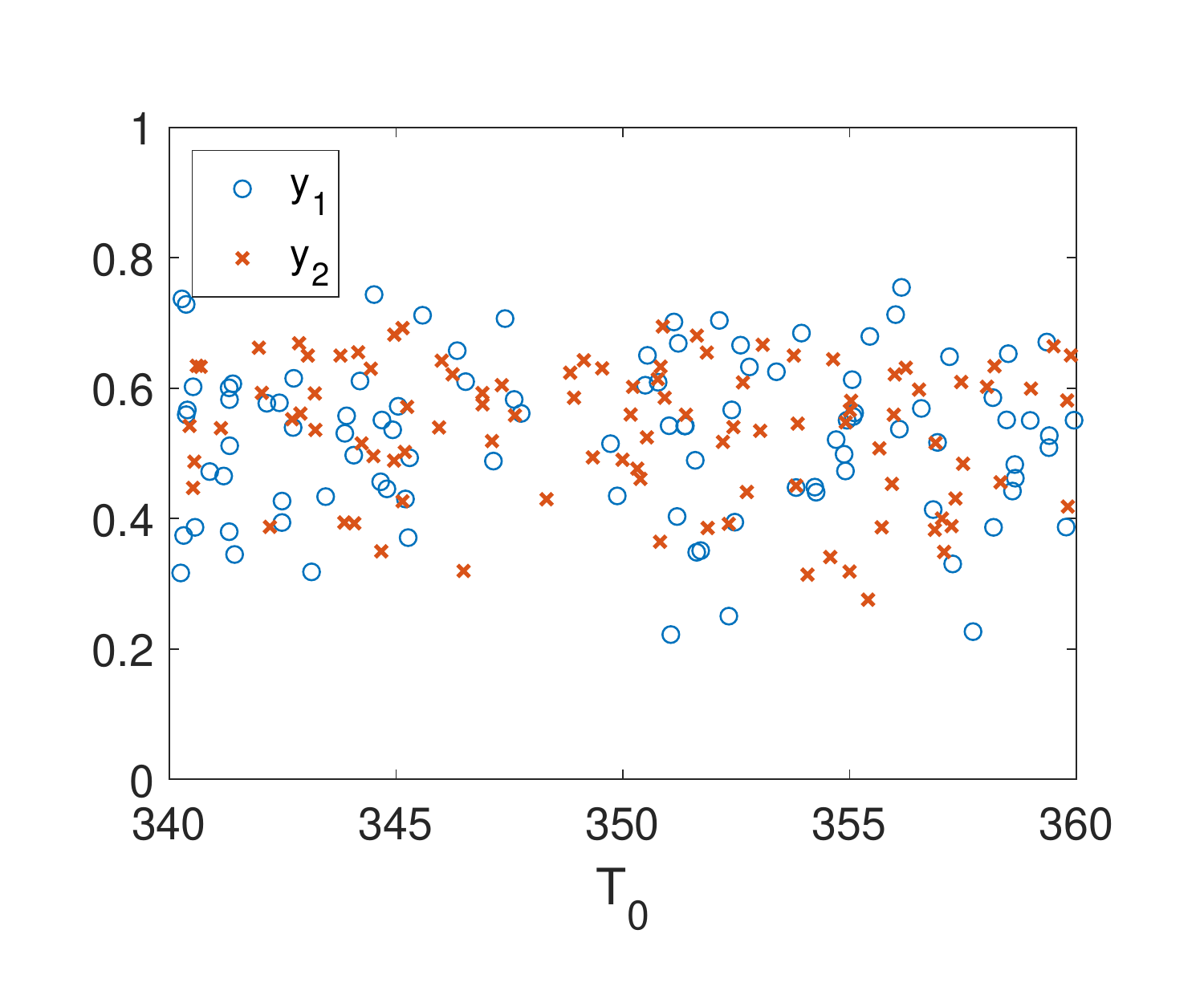}
		\caption{Plots for the piston function. Top left: $ f(x) $; Top right: $ g(x) $; Bottom left: noisy responses versus $ V_0 $; Bottom right: noisy responses versus $ T_0 $.}\label{PistonPlots}
	\end{figure}
	A perturbation on the function, $ g(x) $, is obtained by changing the value of the the spring coefficient from $ k = 2,000 $ to $k = 2,500$ . The range of the function is approximately between $[0.3,0.7]$, so the value of the noise standard deviation is set at $\sigma_\epsilon = 0.05$.
	Figure~\ref{PistonPlots} presents $ f(x) $ and its perturbation, $ g(x) $ along with the noisy datasets. \\
	\textbf{Borehole simulation function}\\
	The borehole function is used to model the flow of water through a borehole~\citep{harper1983} and has been widely used for computer experiments. See, for example,~\cite{Morris-et-al_1993}. The response for this function is the water flow rate in the unit of $ m^3/year $, given by:
	\[
	f(x) = \frac{2\pi T_u(H_u-H_l)}{\ln(r/r_w)\Big(1+\frac{2LT_u}{\ln(r/r_w)r_w^2K_w}+\frac{T_u}{T_l}\Big)},
	\]
	where $ r_w $ is the radius of the borehole ($ m $),
	$ r $ is the radius of the influence ( $ m $),
	$ L $ is the length of the borehole ($ m $),
	$ T_u$ is the transmissivity of the upper aquifer ($ m^2/year $),
	$ T_l $ is the transmissivity of the lower aquifer ($ m^2/year $),
	$ H_u $ is the potentiometric head of the upper aquifer ($ m $),
	$ H_l $ is the potentiometric head of the lower aquifer ($ m $),
	and $ K_w $ is the hydraulic conductivity of the borehole ($ m/year $).
	\begin{figure}
		\centering
		\includegraphics[width = 2.5in]{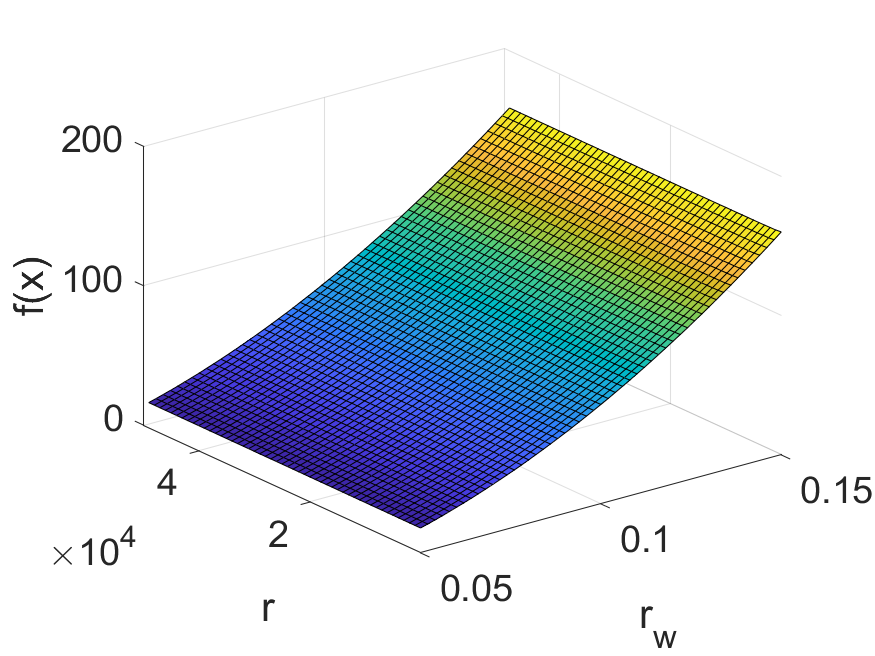}\qquad
		\includegraphics[width = 2.5in]{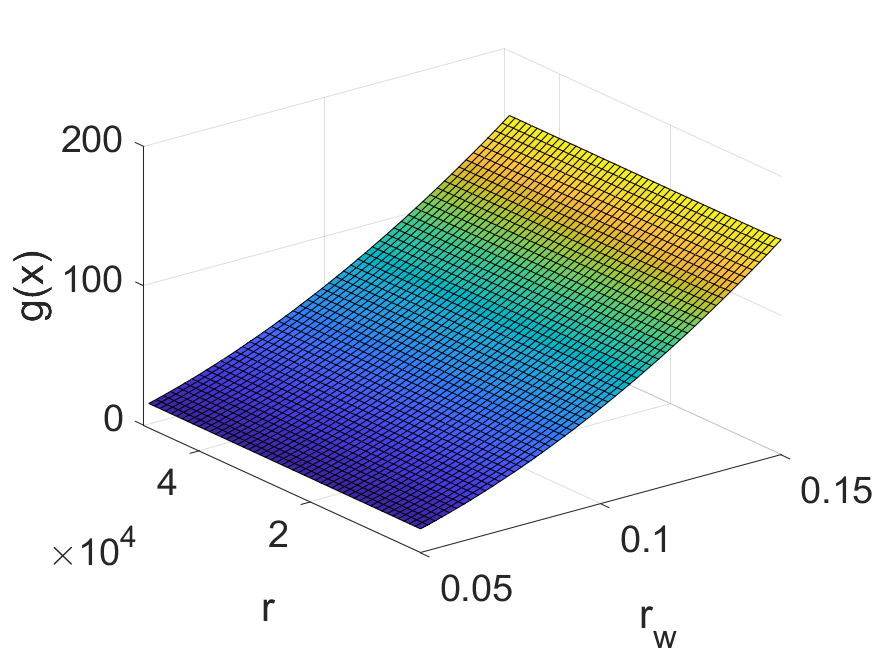}\\
		\includegraphics[width = 2.75in]{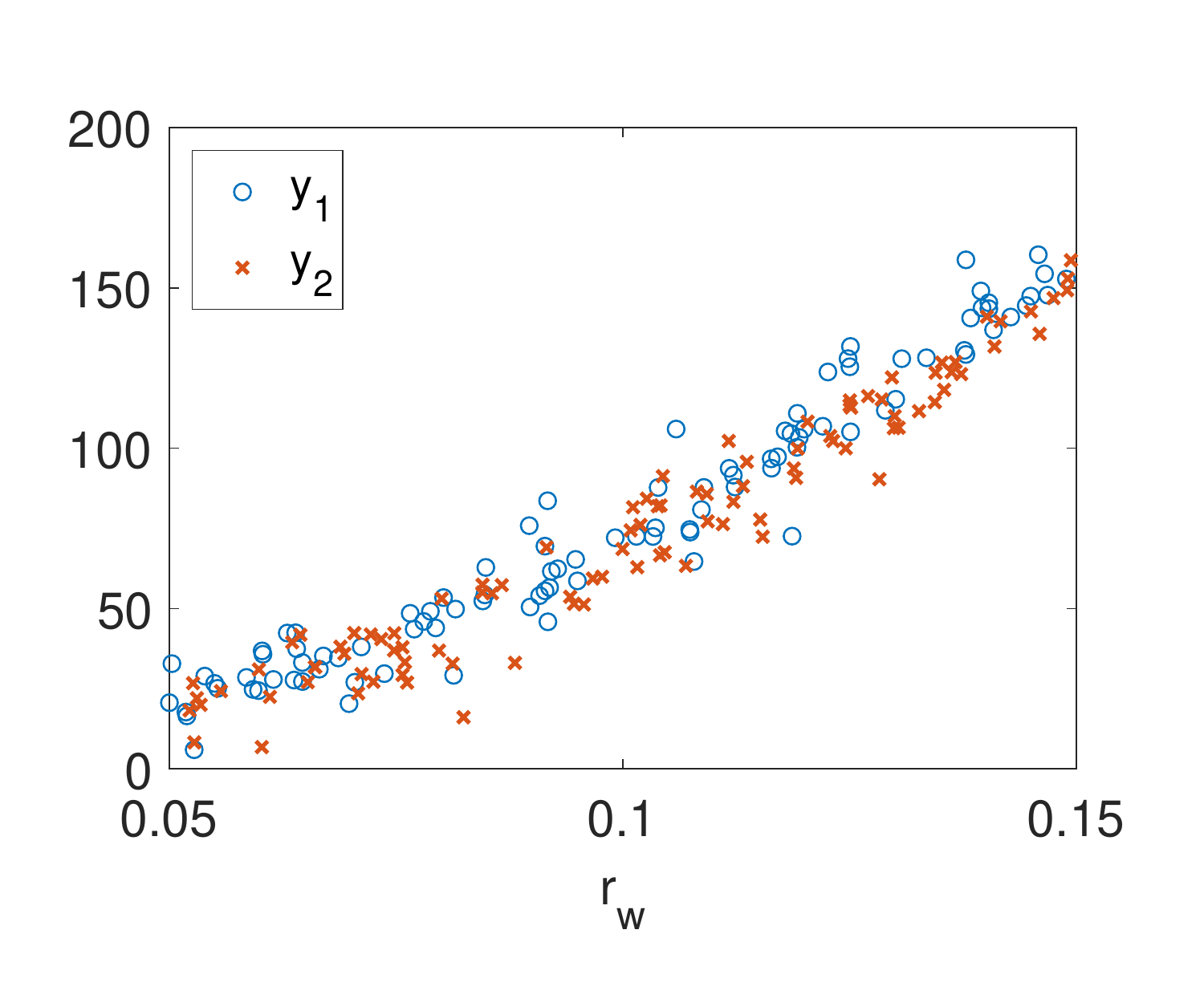}\qquad
		\includegraphics[width = 2.75in]{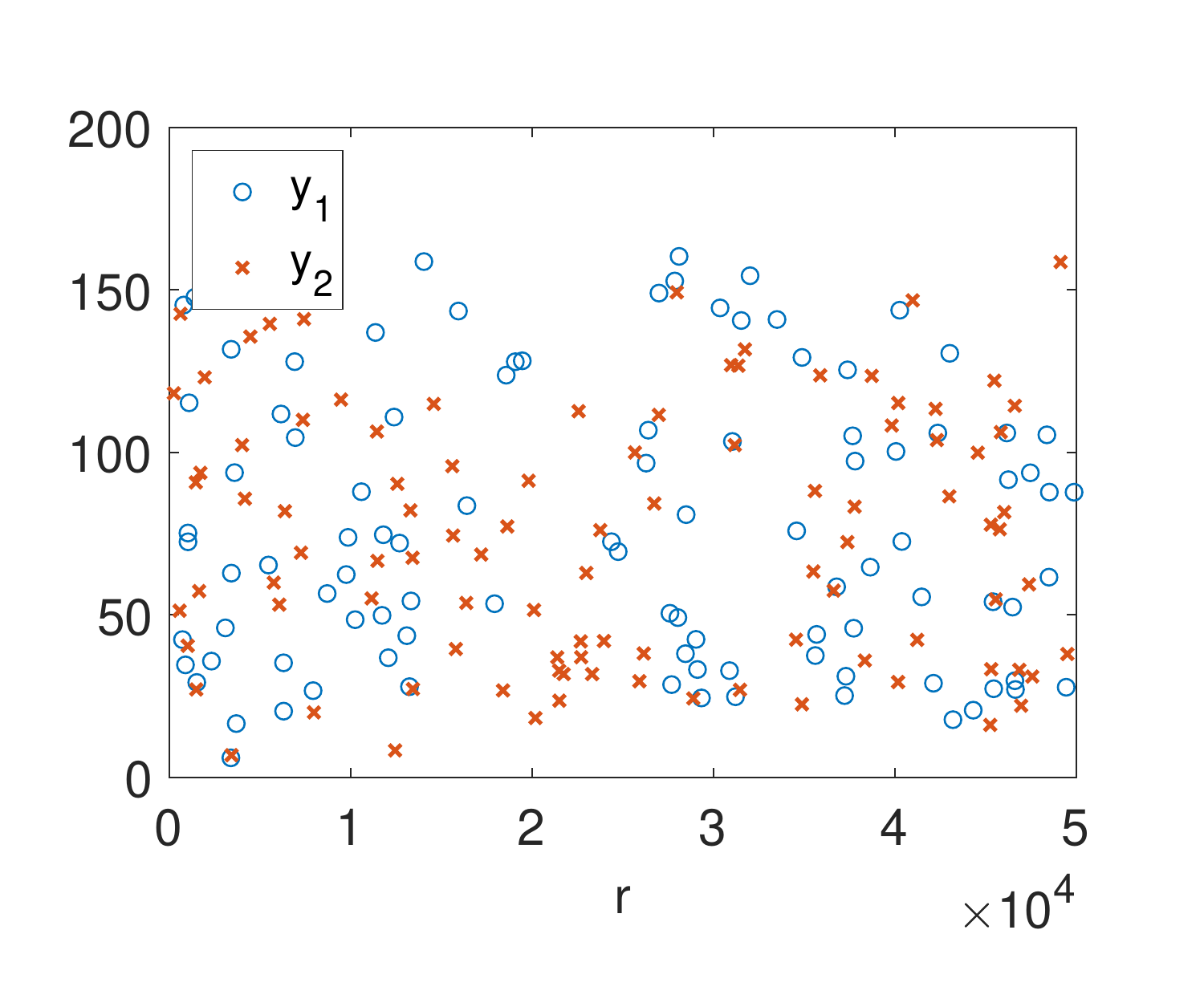}
		\caption{Plots for the borehole function. Top left: $ f(x) $; Top right: $ g(x) $; Bottom left: noisy responses versus $ r_w $; Bottom right: noisy responses versus $ r $.}\label{BoreholePlots}
	\end{figure}
	The number of input variables for the borehole function is eight. Again, we only consider two input variables ($ r $ and $ r_w $) while fixing other variables are fixed at $T_u = 78,000$, $H_u=1,050$, $T_l = 84$, $H_l = 760$, $L = 1,400$, $K_w = 11,000$. In this simulation study, a perturbation, $ g(x) $, is obtained by changing the value of $ L $ from 1400 to 1450. The range of this function is approximately between $[0,150]$, so we set the value of the noise standard deviation at $\sigma_\epsilon = 10$. Figure~\ref{BoreholePlots} show the functions and the noisy data plots.
	
	\pagebreak

\end{document}